\documentclass[preprint]{elsarticle}

\usepackage{tabularx}
\newcolumntype{Y}{>{\centering\arraybackslash}X}
\usepackage{enumitem}
\setlist{
  partopsep=0pt,
  topsep=2pt,
  itemsep=0pt,
  parsep=2pt
}

\usepackage{soul}
\usepackage{wrapfig}
\usepackage{multicol}
\usepackage{multirow}
\usepackage{makecell}
\usepackage{booktabs}
\usepackage{float}
\usepackage{amsthm}
\usepackage{amsmath}
\usepackage{amssymb}
\usepackage{mathtools}
\usepackage{nccmath}
\usepackage{graphicx}
\usepackage{amsfonts}
\usepackage{caption}
\usepackage{mdframed}
\usepackage{listings}
\usepackage[dvipsnames]{xcolor}
\usepackage{courier}
\usepackage{caption}[2015/09/20]
\usepackage{algorithm}
\usepackage[noEnd=true, indLines=true]{algpseudocodex}

\usepackage{xspace}

\definecolor{olivegreen}{HTML}{267F99}
\definecolor{lightbrown}{HTML}{a31515}

\lstdefinelanguage{JavaScript}{
  basicstyle=\ttfamily\scriptsize,
  xleftmargin=1em,
  xrightmargin=1em,
  numbers=none,
  keywords={const, typeof, new, true, false, catch, function, return, null, 
            catch, switch, var, if, in, while, do, else, case, break, import, 
            export, {*}, as, from, this, default, throw, import, class},
  keywordstyle=\color{blue},
  identifierstyle=\color{black},
  sensitive=true,
  comment=[l]{//},
  morecomment=[s]{/*}{*/},
  showstringspaces=false,
  commentstyle=\color{Green}\ttfamily,
  stringstyle=\color{lightbrown}\ttfamily,
  morestring=[b]',
  morestring=[b]",
}

\lstdefinestyle{styleA}{
  language=JavaScript,
  ndkeywords={TitleBar, CarbonContext,Provider, CarbonContextProvider, WindowControls, Avatar, Profile},
  ndkeywordstyle=\color{olivegreen}\bfseries,
  morestring=[s]{=\{}{\}},
  stringstyle=\color{lightbrown}\ttfamily,
}

\usepackage[colorlinks=true, allcolors=blue]{hyperref}

\urldef{\exportsURL}\url{https://nodejs.org/api/modules.html\#modules\_exports}
\urldef{\definePropertyURL}\url{https://developer.mozilla.org/en-US/docs/Web/JavaScript/Reference/Global\_Objects/Object/defineProperty}

\theoremstyle{plain} 
\newtheorem{answer}{AQ}
\newcommand{\React}{\textsc{React}}
\newcommand{\ReactRefactor}{\textsc{React\-Refactor}}
\newcommand{\PropDrilling}{\textit{Prop Drilling}}

\newcommand{\listingScale}{0.80}

\begin{document}
\begin{frontmatter}

\title{Refactoring \React{} Component Hierarchies to Eliminate Prop Drilling}

\author[aueb]{Vangelis Gkinis}
\ead{vangelisgkinis1@gmail.com}
\author[uoi]{Vassilis E. Zafeiris\corref{cor1}}
\ead{bzafiris@uoi.gr}

\address[aueb]{Department of Informatics, Athens University of Economics and Business, Greece}
\address[uoi]{Computer Science and Engineering Department, University of Ioannina, Greece}

\cortext[cor1]{Corresponding author. Address: University of Ioannina, 451 10, Ioannina, Greece.}

\journal{Elsevier Journal}

\begin{abstract}
In \React{} front-end development,
\PropDrilling{} is the practice of propagating data
through component properties across multiple levels of the component hierarchy.
Despite being discouraged by \React{} documentation
and characterized as a code smell by recent research,
little is known about its prevalence, complexity, and potential for automated refactoring.
In this work, we propose a static analysis method
for identifying \PropDrilling{} instances in a \React{} codebase
and eliminating them using two automated refactoring strategies
based on \emph{Context API} and \emph{Component Composition}.
The proposed method is implemented as a Node.js command line tool, \ReactRefactor{},
and empirically evaluated on a benchmark dataset of open-source \React{} applications.
The main findings of the empirical evaluation indicate
(a) the frequent occurrence of prop drillings across benchmark projects, 
irrespective of project size; 
(b) their generally low to moderate complexity in terms of data propagation
depth, contrasted with the more prevalent complexity arising from the simultaneous
forwarding of multiple properties along the same path; and, 
(c) the potential for automated elimination of a substantial share (76.3\%) of prop drillings,
primarily through refactoring to the Context API.
\end{abstract}


\end{frontmatter}


\newcommand{\rephraseStart}{\color{blue}}
\newcommand{\rephraseEnd}{\color{black}}

\newcommand{\considerStart}{\color{red}}
\newcommand{\considerEnd}{\color{black}}

\newcommand{\ctext}[2]{
  \begingroup
  \sethlcolor{#1}
  \hl{#2}%
  \endgroup
}

\newcommand{\silence}[1]{}
\newcommand{\forcepagebreak}{\clearpage \newpage}

\newcommand{\highlight}[1]{\ctext{yellow}{#1}}

\newcommand{\code}[1]{\lstinline!#1!}
\newcommand{\tool}[1]{\textsc{#1}}
\newcommand{\proj}[1]{\textsc{#1}}
\newcommand{\render}{\code{render()}}
\newcommand{\Component}{\code{Component}}
\newcommand{\Editor}{\code{Editor}}
\newcommand{\WindowControls}{\code{WindowControls}}
\newcommand{\Carbon}{\code{Carbon}}
\newcommand{\TitleBar}{\code{TitleBar}}
\newcommand{\titleBar}{\code{titleBar}}
\newcommand{\updateTitleBar}{\code{updateTitleBar}}
\newcommand{\onTitleBarChange}{\code{onTitleBarChange}}
\newcommand{\lightProp}{\code{light}}
\newcommand{\valueProp}{\code{value}}
\newcommand{\onChange}{\code{onChange}}
\newcommand{\children}{\code{children}}
\newcommand{\CarbonContext}{\code{CarbonContext}}
\newcommand{\CarbonCtxProvider}{\code{CarbonContext.Provider}}
\newcommand{\CarbonCtxProviderWrap}{\code{CarbonContextProvider}}
\newcommand{\ContextProvider}{\code{Context.Provider}}
\newcommand{\createContext}{\code{createContext()}}
\newcommand{\useContext}{\code{useContext()}}

\newcommand{\FunctionDeclaration}{\code{FunctionDeclaration}}
\newcommand{\ClassDeclaration}{\code{ClassDeclaration}}
\newcommand{\VariableDeclarator}{\code{VariableDeclarator}}
\newcommand{\extractComponent}{\code{extractComponent()}}
\newcommand{\PureComponent}{\code{PureComponent}}
\newcommand{\extractFQN}{\code{extractFQN()}}
\newcommand{\extractProps}{\code{extractProps()}}
\newcommand{\extractUses}{\code{extractComponentUses()}}
\newcommand{\resolveEdges}{\code{resolveEdges()}}
\newcommand{\resolvePropForwardings}{\code{resolvePropForwardings()}}
\newcommand{\identifyPropDrillings}{\code{identifyPropDrillings()}}
\newcommand{\searchPropDrillings}{\code{searchPropDrillings()}}
\newcommand{\getMatchingPF}{\code{getMatchingPF()}}

\newcommand{\Actx}{$A_{ctx}$}
\newcommand{\pure}[1]{\bar{#1}}
\newcommand{\nonpure}[1]{\bar{#1}'}
\newcommand{\theading}[1]{#1}

\newcommand{\valpct}[2]{#1 (#2\%)}
\newcommand{\pct}[1]{#1\%}
\newcommand{\avg}[2]{#1 (#2)}
\newcommand{\emptyVal}{--}

\newcommand{\colhead}[2]{\multicolumn{1}{|p{#1}|}{\raggedright\textbf{#2}}}
\newcommand{\Ca}[0]{$C_a$}
\newcommand{\Cb}[0]{$C_b$}
\newcommand{\Ci}[0]{$C_i$}
\newcommand{\Cs}[0]{$C_s$}
\newcommand{\Ct}[0]{$C_t$}
\newcommand{\Pa}[0]{$P_a$}
\newcommand{\Pb}[0]{$P_b$}
\newcommand{\pj}[0]{$p_j$}
\newcommand{\parent}[2]{$#1\rightarrow{}#2$}
\newcommand{\notparent}[2]{$#1\nrightarrow #2$}
\newcommand{\seqlength}[1]{\left\lvert #1 \right\rvert}
\newcommand{\breadth}[1]{$\vert$#1$\vert$}


\newcommand{\carbon}{\proj{Carbon}}

\newcommand{\javascript}{JavaScript\xspace}
\newcommand{\jscodeshift}{jscodeshift\xspace}

\section{Introduction} 
\label{introduction}

\React{} is a library for implementing dynamic and interactive 
web user interfaces in JavaScript (JS) and TypeScript (TS)~\cite{reactDev}.
It supports modular and declarative development
of front-end applications for the Web Browser platform.
A \React{} application implements the screens
and the corresponding front-end logic of a web application.
Each screen is structured in terms of \emph{web components}
or simply \emph{components}, in \React{} terminology,
which encapsulate fragments of the screen presentation and behaviour.
Components are implemented as functions or classes
and declare their display view in JSX,
a JS/TS extension with support for embedded XML-like markup syntax.
\React{} favors component \emph{composition} 
and allows for specification of complex view structures
in terms of hierarchically nested tags
corresponding to standard HTML elements and 
elements representing other \React{} components.

At runtime, the components comprising a screen are \emph{rendered}
to recursively generate a tree representation of the view structure,
which is commonly referred to as \emph{Virtual DOM} or \emph{render tree}.
Notice that, component rendering involves invocation of functional components
or invocation of the \render{} method for class components.
The Virtual DOM is, initially, constructed on page load,
and the actual DOM is updated by harnessing content
from \emph{host components}, residing typically 
at the lower levels of the tree~\cite{reactFiber}.
The arrival of events that update the UI state 
(e.g. user interactions, AJAX calls)
trigger the generation of a new Virtual DOM
through a process called \emph{reconciliation}.
During reconciliation the new Virtual DOM is constructed
through traversal of the current Virtual DOM 
and \emph{re-rendering} of components that require updates 
or reuse of unchanged portions of the tree.
Changes between the current and the new Virtual DOM
are, then, applied to the actual DOM (commit phase),
minimizing, thus, costly DOM manipulations and improving performance.

On a technical basis, the reconciliation process
is triggered by a \emph{state} update,
i.e. an update to the value of a component's \emph{state} variable
somewhere in the render tree.
In case that multiple state updates occur within the same event handler,
\React{} batches them and typically performs a single reconciliation pass.
State variables and their respective state update (or ``setter'') functions
are declared at the component level,
through appropriate \React{} system calls,
called \React{} \emph{hooks}.
State variables are distinct from locally declared variables
in that they retain their values across component re-renders.
Separate copies of state variables
are associated with each component position in the render tree.

State management is one of the most challenging aspects of \React{} development,
involving both the handling of state updates
and the propagation of data across the component hierarchy.
State update deals with the flow of data from external sources
(e.g. user input, network requests) towards the render tree.
Developers must decide where state should be maintained 
within the component hierarchy and 
how it should be structured to avoid unnecessary re-renders. 
On the other hand, data propagation concerns
the flow of information within the render tree
and occurs during the initial rendering of the application,
as well as on subsequent re-renders.
\React{} follows a unidirectional data-flow model
where higher-level components forward to descendant components
data that are required for their rendering.
A standard approach for data propagation 
employs component properties or \emph{props};
a component receives the required data 
through value assignments to its props (prop passings)
in the returned JSX of the parent component.
In case that the data owner resides higher in the component hierarchy,
data are forwarded through successive prop passings
across intermediate components.
\React{}, also, provides the \emph{Context API} 
that allows data to be shared across entire subtrees without 
explicitly passing properties through intermediate components.

The practice of propagating data through component properties 
across multiple levels of the component hierarchy
is known as \PropDrilling{}.
According to the official \React{} documentation,
\PropDrilling{} should be avoided where possible,
in favor of alternatives, such as the Context API~\cite{reactDevCtx}.
More recently, research on the maintainability of \React{} codebases
classifies \PropDrilling{} as a code smell
due to the difficulty in tracing the origin and uses of forwarded property values,
especially in large component hierarchies,
and the tight coupling introduced among participating components
~\cite{Ferreira2023,Ferreira24,HwangKRLPJJS26}.

In this work, we propose a static analysis method 
for identifying \PropDrilling{} instances  
and eliminating them using two automated refactoring strategies
based on the \emph{Context API} and \emph{Component Composition},
native \React{} capabilities that facilitate the implementation
of adaptable and reusable components.
The identification of prop drilings is based 
on a graph representation of the \React{} codebase --
the Component Hierarchy Graph (CHG) --
that statically approximates the parent-child relationships among components
and the flow of data due to prop passings and prop forwardings.
A set of refactoring preconditions determine
whether a \PropDrilling{} instance is refactorable
and which refactoring strategies are applicable.
Finally, we specify source code transformation procedures
for eliminating a \PropDrilling{} instance through either refactoring strategy.

The proposed method has been implemented
as a Node.js command line tool, \ReactRefactor{},
that analyzes efficiently large \React{} codebases 
and applies refactorings under user guidance.
We employed \ReactRefactor{} to empirically evaluate our method
on a benchmark dataset of 15 open-source \React{} projects,
in terms of its applicability, effectiveness and practicality.
Among the main findings of the empirically study are
(a) the frequent occurrence of prop drillings across benchmark projects,
irrespective of project size; 
(b) their generally low to moderate complexity in terms of data propagation depth, 
contrasted with the more prevalent complexity due to the simultaneous forwarding
of multiple properties along the same path; and,
(c) the potential for automated elimination of a substantial share (76.3\%) of prop drillings,
primarily through refactoring to the Context API.

To the best of our knowledge this is the first work 
that studies the automation of a large-scale refactoring in a \React{} codebase.
Moreover, it is among the few studies 
that investigate the prevalence of code smells in real-world \React{} applications
and the first to focus specifically on \PropDrilling{}.
Thus, the main contributions of this paper are as follows:
\begin{itemize}
  \item a static analysis method for identifying and automatically eliminating 
    \PropDrilling{} instances through refactoring,
  \item an empirical study of the prevalence 
      and complexity of prop drillings in real-world applications,
      as well as their potential for automated elimination through refactoring,
  \item an open-source implementation of the proposed method, \ReactRefactor{},
   for analyzing component hierarchies and refactoring prop drillings.
\end{itemize}

The rest of the paper is organized as follows. 
Section~\ref{sect:relatedwork} presents related work and 
Section~\ref{sect:prop-drilling} provides 
a specification of the \PropDrilling{} code smell
and its implications on code maintainability.
Moreover, in Section~\ref{sect:refactoring},
we specify our method for identifying and refactoring
\PropDrilling{} instances in a \React{} codebase.
Section~\ref{sect:evaluation} presents the experimental setup
and the results of our empirical study,
while Section~\ref{sect:ttv} discusses potential threats to its validity. 
Finally, the paper is concluded in Section~\ref{sect:conclusions}.

\section{Related Work} 
\label{sect:relatedwork}

Our work is, primarily, relevant to two lines of research 
within the broader area of Software Maintenance and, specifically,
(a) definition of code smells or anti-patterns 
and their automated detection in a given codebase, and
(b) refactoring automation, involving research topics
on automated detection of refactoring opportunities
and automated application of refactorings 
through behaviour-preserving code transformations.
Given the volume of research contributions to these topics
for at least three decades, 
we narrow down the scope of this section's literature review
to works relevant to the JS/TS web application platform.
A reader seeking a more general perspective on this area
can refer to the seminal work of Martin Fowler~\cite{Fowler99} 
and its 2nd edition specialized to the JS language~\cite{Fowler18},
as well as to recent surveys on code smells and refactoring,
e.g.~\cite{lacerda2020},~\cite{abuhassan2021},~\cite{thakur2026}.

A first approach in the area of code smell detection 
in JS code is \tool{JSNose}~\cite{Fard2013}.
The authors propose a catalog of 13 code smells for web applications
and a technique, implemented in the \tool{JSNose} tool,
that combines static and dynamic analysis for their detection.
Later,~\citet{Gong15} document 28 common pitfalls
related to code quality violations in JS code,
while~\citet{Xiao2015} propose a method for identification
of 6 types of performance-related code smells in JS programs
and suggests potential refactorings for their elimination.
~\citet{Saboury2017} focuses on server-side JS applications
and studies the presence of 12 types of code smells in a dataset of 5 open-source libraries.
Moreover, the authors conducted a developer survey 
to understand developer perception on the severity of code smells 
and their impact on maintainability.
An extension to this work analyzed the release history of 15 JS applications
and investigated the correlation of code smells with the error-proneness of related source files,
as well as the survival of code smells in the evolution history of analyzed projects~\cite{Johannes2019}.

More recent works, studied code smells from the perspectives
of concurrency, security and dependency management.
Specifically,~\citet{Turcotte2022} focus on promise-based JS code
and propose 8 anti-patterns related to the use of the Promise API
and async/await constructs.
The study, also, introduces \tool{DrAsync},
a dynamic analysis tool for detection and visualization of promise lifetimes,
and provides an empirical study on the presence of anti-patterns
and their potential for refactoring.
Concerning the area of security,
~\cite{Kambhampati2024} proposes 24 JS security smells,
affecting both client and server side code, 
and extends \tool{JSNose} to detect them.
The code smells are, also, mapped
to CWE and OWASP Top 10 categorizations of security risks.
Finally,~\cite{Jafari2022} introduces a catalog of dependency smells in \code{npm} projects,
based on responses from a developer survey,
and investigates their prevalence 
through a large scale empirical study on 1,146 popular JS projects.

Research on framework-specific code smells remains 
relatively scarce despite the widespread adoption of front-end frameworks.
Regarding the \React{} framework,
~\citet{Ferreira2023} proposed a catalog of 12 code smells and a tool, \tool{ReactSniffer}, 
for their automated detection in a given codebase.
Moreover, they conducted an empirical study 
on the prevalence of code smells in real-world applications
excluding, however, \PropDrilling{} 
due to limitations of the \tool{ReactSniffer} tool.
According to the authors, 
the detection of prop drillings (and duplicated components)
requires sophisticated static analysis 
and their plan is to support it in future versions of the tool.
More recently,~\citet{Nunes2025} extended this work
with a set of 6 TypeScript-related code smells
and studied their presence in open-source projects through \tool{SniffTSX}.
The study included also a developer survey on their severity.

In parallel to the study of code smells, 
several research efforts have focused on methods
and tools for refactoring automation.
Early approaches explored basic refactorings,
with particular emphasis on the challenges 
introduced by the dynamic features of the JS language.
\citet{Feldthaus2011} propose a static analysis framework 
for specifying tool-supported refactorings
and showcase their approach through the implementation 
of rename/encapsulate property and extract module refactorings.
Renaming JS object properties through refactoring 
has been further studied in \cite{Feldthaus2013}.
The authors employ static analysis 
for finding related identifiers to rename
and implement their approach as an \tool{Eclipse} plugin.
\citet{Jensen2012} focus on the notorious \code{eval} call
for dynamic parsing and execution of JS code strings
and propose a method, based on data flow analysis,
to refactor \code{eval} invocations to equivalent JS code.
The method is implemented in the \tool{Unevalizer} framework 
and empirically evaluated in production web sites,
confirming its capability to eliminate typical cases of \code{eval} usage.

Later approaches focused on refactorings for modernization of legacy code
with emphasis on the adoption of language features standardized in ECMAScript 2015 (ES6)
that handle concerns related to concurrency, encapsulation and modularity.
\citet{Ying2013} introduce an approach 
to convert legacy XML-based AJAX applications to JSON-based 
and evaluate the efficiency gains with the \tool{XtoJ} research prototype.
The migration of asynchronous code to the Promise API -- standardized in ES6 --
is studied in~\cite{Brodu2015,Gallaba2017}.
~\citet{Gallaba2017}, also, conduct an empirical evaluation
of their automated refactoring method with respect to its effectiveness and practicality.
The evaluation is based on the \tool{PromisesLand} research prototype.
\citet{Gokhale2021} propose a refactoring method
for automated migration of calls to synchronous APIs to their asynchronous counterparts.
The method has been implemented in the \tool{Desynchronizer} tool for evaluation purposes.

Several studies investigated the migration to ES6 classes and modules,
to support encapsulation and modularity through modern, high-level language constructs.
Specifically,~\citet{Rostami16} introduce \tool{JSDeodorant}, 
an Eclipse plugin for detecting function constructors that emulate classes
and represent opportunities for refactoring to ES6 classes.
\cite{SilvaJSEP17}, also, focus on the identification of class emulations, 
proposing a method and tool, called \tool{JSClassFinder}.
Their work, also, involves an empirical study
on the presence of class/subclass emulations in legacy ES5 code,
and the accuracy of the proposed method.
Regarding the actual refactoring to ES6 classes,
\citet{SilvaICSR2017} specify a set of rules to transform ES5 class emulations 
to class definitions using the ES6 syntax.
Finally,~\cite{Paltoglou2018,Paltoglou2021} introduce a fully-automated method
for refactoring of legacy ES5 code to ES6 modules,
with fine-grained reuse of module contents
through the \code{named import/export} language constructs. 

Research on refactoring framework-specific JS/TS code
has only recently gained attention, primarily focusing on \React{}.
\citet{Ferreira24} introduce a catalog of 69 refactoring operations for \React{} applications,
derived through manual inspection of refactoring commits in open-source projects.
Moreover,~\citet{Lobo2026} describe an extension of the \tool{ReactSniffer} tool~\cite{Ferreira2023} 
and its integration into the Visual Studio Code IDE.
Their work focuses on the automation of seven refactoring operations,
included in the catalog of~\citet{Ferreira24},
but does not specify a method for identification of refactoring opportunities
or the required code transformations for each refactoring.

In this paper we provide an in-depth study of the \PropDrilling{} code smell.
We formally describe its structure
and identify inherent properties that characterize its complexity.
Furthermore, we investigate the prevalence 
and complexity of prop drillings in real-world \React{} applications,
complementing the work of~\citet{Ferreira2023}
and advancing our understanding of data propagation 
in \React{} component hierarchies.
The implemented research prototype, \tool{ReactRefactor}, 
integrates advanced static analysis capabilities 
for recovering component hierarchies and 
tracking data flow across components due to prop passing.
Regarding refactoring automation, 
our work is the first to study a complex refactoring in \React{} code.
Compared with~\cite{Lobo2026},
we additionally specify refactoring preconditions 
to preclude the suggestion of unsafe refactorings
and describe the steps of the code transformation.
The proposed refactoring is fully implemented in \tool{ReactRefactor}
for evaluation purposes.

\section{The Prop Drilling Code Smell} 
\label{sect:prop-drilling}

In this section,
we introduce the core concepts, terminology and definitions 
related to the \PropDrilling{} code smell.
Moreover, we discuss alternative designs
for the elimination of \PropDrilling{} instances through refactoring
with emphasis on approaches relying on native capabilities of the \React{} platform.
Specifically, we demonstrate how the \emph{Context API} and \emph{Component Composition}  
enable data propagation across multiple levels of the component hierarchy,
thereby avoiding the need for \PropDrilling{}.

The concepts and approaches presented in this section
are clarified using a motivating example from the \proj{Carbon} project,
which is part of the benchmark dataset used in the empirical study of Section~\ref{sect:evaluation}.
The \proj{Carbon} project implements a code editor
that generates beautiful images of code snippets 
and publishes them to social media.
It supports syntax highlighting
and rich customization capabilities for the generated images.

\begin{figure}[htb] 
    \centering
    \includegraphics[scale=0.65]{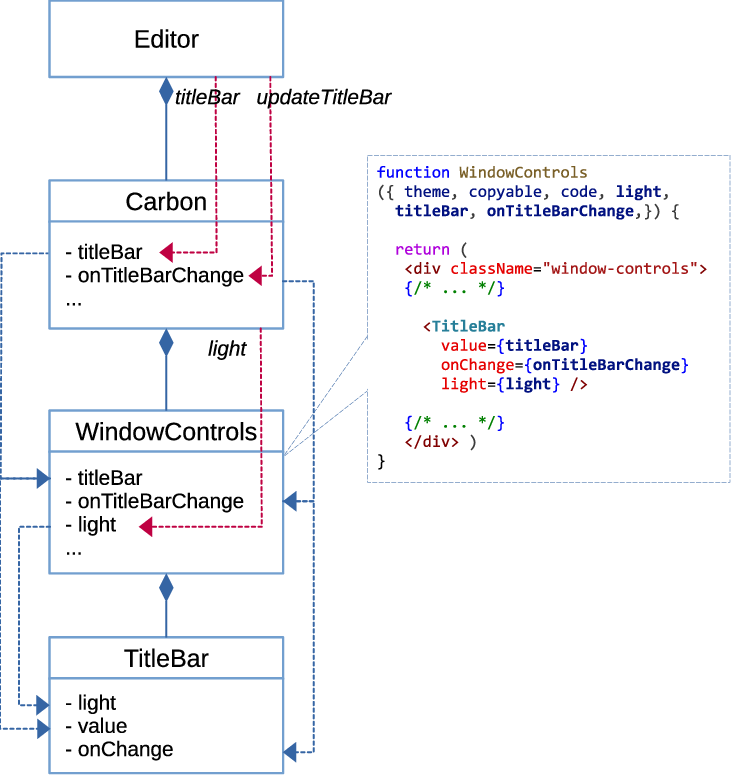}
    \captionsetup{justification=centering}
    \caption{Prop Drilling example from the \proj{Carbon} project.}
    \label{fig:prop-dril-example}
\end{figure}

Figure~\ref{fig:prop-dril-example} presents 
a part of the project's component hierarchy, 
comprising core components of the application's UI.
Specifically, \Editor{} represents the top-level component of the UI,
while \Carbon{} shows the highlighted code snippet 
in an area that resembles a terminal emulator window.
\WindowControls{} implements the title bar area of the window
and \TitleBar{} presents the actual title.
As will be further discussed in the rest of this section,
\proj{Carbon} developers resorted to the \PropDrilling{} approach,
in order to propagate configuration information 
from the \Editor{} component to the deeply nested \TitleBar{} component.
A detailed description of the identified \PropDrilling{} instances
will be provided after a formal definition of the \PropDrilling{} concept.

\subsection{Core concepts and definitions}
\label{sect:definitions}

Let \Ca{}, \Cb{} be two components with a parent-child relationship,
i.e. \Ca{} has the role of the parent and references \Cb{} 
in the JSX markup returned by its function constructor.
We will, henceforth, refer to this relationship
with the notation \parent{C_a}{C_b}.
Moreover, let \Pa{}, \Pb{} be the set of properties
implicitly or explicitly declared in their respective function constructors.
For brevity reasons and without loss of generality,
the specification of the proposed method will mostly focus on functional components.
This, also, keeps up with the official \React{} documentation
that advises the use of functions for component implementation~\cite{reactDev}.
Nevertheless, the implementation of our method
and empirical study results 
take into account both class and function components.

A \emph{Prop Passing}, in \React{} jargon,
is a value assignment to a child component's property,
in order to configure its rendering.
We represent a \emph{Prop Passing} as a triple 
$PP_{a,b,j} = (C_a, C_b, p_j)$, 
where \parent{C_a}{C_b} and
\Ca{} sets the value of property $p_j \in P_b$.
Prop passings are depicted in Figure~\ref{fig:prop-dril-example}
as red arrows that originate from a component
and target a child component's property.
For instance, \code{titleBar} and \code{updateTitleBar} arrows
from \Editor{} component represent local variables,
whose values are assigned to \code{titleBar}
and \code{onTitleBarChange} properties of \Carbon{}, respectively.

We define \emph{Prop Forwarding}, 
as a special case of prop passing,
where the value assigned to $p_j$ 
is a reference to a property $p_i \in P_a$ of the parent component \Ca{}
and the value of $p_i$ is not defined in the implementation of \Ca{}. 
Notice that although component properties are considered immutable,
a component may define, during rendering, 
a property to a new value that is not persisted in future re-renders.
A \emph{Prop Forwarding} is represented as a quadruple:
\[PF_{a,b,i,j} = (C_a, C_b, p_i,\ p_j),\ C_a\rightarrow{}C_b \land p_i \in P_a \land p_j \in P_b \land p_j \coloneq p_i\]
We will refer to $p_i$, $p_j$ as source and target properties, respectively.
In case that the source property $p_i$ is not referenced in any other location
in the implementation of \Ca{},
$PF_{a,b,i,j}$ will be referred to as \emph{Pure Prop Forwarding}
and will be represented with the notation $\overline{PF}_{a,b,i,j}$.

In Figure~\ref{fig:prop-dril-example},
components are drawn with UML class notation,
while their properties are represented as class attributes.
The composition relationship denotes 
the parent-child relationship among components,
e.g. \Editor{} is the parent of \Carbon{} and so on.
Moreover, prop forwardings are presented as blue arrows,
that associate pairs of properties among parent and child components.
For instance, the \onTitleBarChange{} property of \Carbon{}
is forwarded to a \TitleBar{} property with the same name.
Notice that all prop forwardings in the motivating example are pure.
The right part of the figure depicts a code snippet
from the implementation of \WindowControls{}
that highlights its pure prop forwardings.
Specifically, the \lightProp{}, \titleBar{} and \onTitleBarChange{} 
properties of the component are directly forwarded
as values of the \lightProp{}, \valueProp{} and \onChange{}
properties of \TitleBar{}, respectively.

The \PropDrilling{} concept can, now, be defined
on the basis of \emph{Prop Passing} and \emph{Prop Forwarding}.
Let \Cs{}, \Ct{} be a pair of components, such that $C_s \neq C_t$.
A \emph{Prop Drilling} from \Cs{} to \Ct{}
with respect to property $p_j \in P_t$
is defined as a sequence of prop passings $PD_{s,t,j}=(PP_0, PP_1, \ldots, PP_n)$.
We will refer to \Cs{}, \Ct{} and \pj{} as 
\emph{source} component, \emph{target} component 
and \emph{target} property
and to the rest of the components 
that participate in $PD_{s,t,j}$ as \emph{intermediate} components.
The source, target and intermediate components of a prop drilling,
will be referred to as \emph{participant} components 
or, in short, \emph{participants}.
The sequence of participants, in the order that they are encountered in $PD_{s,t,j}$,
is the prop drilling path, i.e $path(PD_{s,t,j})=(C_s, C_0, \ldots, C_{n-1}, C_t)$.

A prop drilling $PD_{s,t,j}=(PP_0, PP_1, \ldots, PP_n)$ has the following properties:

\begin{itemize}
    \item \notparent{C_s}{C_t}, i.e., the source and target components do not share
    a parent-child relationship,
    \item $\seqlength{path(PD_{s,t,j})} \geq 3$, i.e., it includes at least
    one intermediate component,
    \item $PP_0\equiv PP_{s, 0, k}$, the first element is a prop passing
    from the source component $C_s$ towards some component $C_0$ 
    with respect to a property $p_k \in P_0$,
    \item $PP_n\equiv PF_{n-1, t, l, j}$, the last element is a prop forwarding
    from a component $C_{n-1}$ to the target component $C_t$
    and its property $p_j \in P_t$,
    \item $PP_i \equiv PF_{i-1, i, r, s}, i \in [1, n-1]$, any other element is, also,
    a prop forwarding from $C_{i-1}$ to $C_{i}$,
    \item given any pair of consecutive elements $(PP_{i-1}, PP_i)$,
    it holds that $PP_{i-1}\equiv PF_{i-2, i-1, m, r} \land PP_i \equiv PF_{i-1, i, r, s}$,
    i.e., the target property of a prop forwarding 
    matches the source property of the next prop forwarding in the sequence.
\end{itemize}

Thus, within a prop drilling instance $PD_{s,t,j}$,
a value defined in the source component \Cs{} 
is forwarded through properties of intermediate components
and is, finally, assigned to the property \pj{} of the target component \Ct{}.
A prop drilling is terminated on the discontinue of prop forwardings
or in case that a prop forwarding alters the forwarded property value.


\emph{Prop Drilling Depth.} The impact of \PropDrilling{} on code maintainability
is affected by the number of intermediate components
in \PropDrilling{} instances.
The higher the number of intermediates,
the more components are unnecessarily coupled
to the source and target components
and affected by their data dependencies.
Towards this direction,
we assess the severity of a \PropDrilling{} instance
with the \emph{depth} metric
that is equal to the number of intermediates
or equivalently to the number of prop forwardings
in the prop drilling sequence
(each intermediate starts a new prop forwarding).
By definition, the minimum depth value is 1,
since a prop drilling includes at least one prop forwarding.

\emph{Compound Prop Drilling.} Besides depth, 
a further indication of the coupling level incurred due to \PropDrilling{},
is the number of properties that are simultaneously forwarded
along the same \emph{path}.
Let $PD_i \equiv PD_{s,t,j_i}, i \in [1, N]$ be a set of prop drillings,
sharing the same path and
targeting properties from the set $J=\bigcup{\{j_i\}}, i \in [1, N]$.
The set $CPD_{s,t, J} = \bigcup \{PD_{s,t,j_i}\}, i \in [1, N]$
will be referred to as \emph{Compound Prop Drilling}
from source component \Cs{} to target \Ct{}
with respect to the set of properties $J\subseteq P_t$.
The cardinality of the property set $J$,
will be referred to as \emph{breadth} of $CPD_{s,t, J}$.
The occurrence of a compound prop drilling 
denotes a stronger coupling among participant components
that increases with the breadth of the prop drilling.
The more properties forwarded,
the higher chance for further changes (deletions, renames)
with rippling changes along the prop drilling path.
It also reveals an intent for data sharing among source
and target components and not incidental data forwarding.

Figure~\ref{fig:prop-dril-example} illustrates
three prop drilling instances
identified in the \proj{Carbon} project.
All prop drillings target the properties 
of the \TitleBar{} component.
The source component in two of them is \Editor{},
which passes the \titleBar{} and \updateTitleBar{} local variables
to respective properties of \Carbon{} (indicated by the red arrow endings).
These properties are, then, forwarded
to the \valueProp{} and \onChange{} properties of \TitleBar{},
in a series of pure prop forwardings indicated by the blue arrows.
The \emph{depth} of both prop drillings is two,
while they form a \emph{Compound Prop Drilling} of breadth two.
Concerning the third prop drilling,
it starts from the \Carbon{} component with a prop passing
towards the \lightProp{} property of \WindowControls{} component.
The property is, finally, provided to \TitleBar{},
through a pure prop forwarding.
Note that the last prop drilling 
can be embodied to the compound prop drilling,
assuming that \Carbon{} is its source.
In this case the breadth of the resulting prop drilling is three.

\subsection{Alternative Designs}

\begin{figure*}[!htb] 
    \centering
    \includegraphics[scale=0.75]{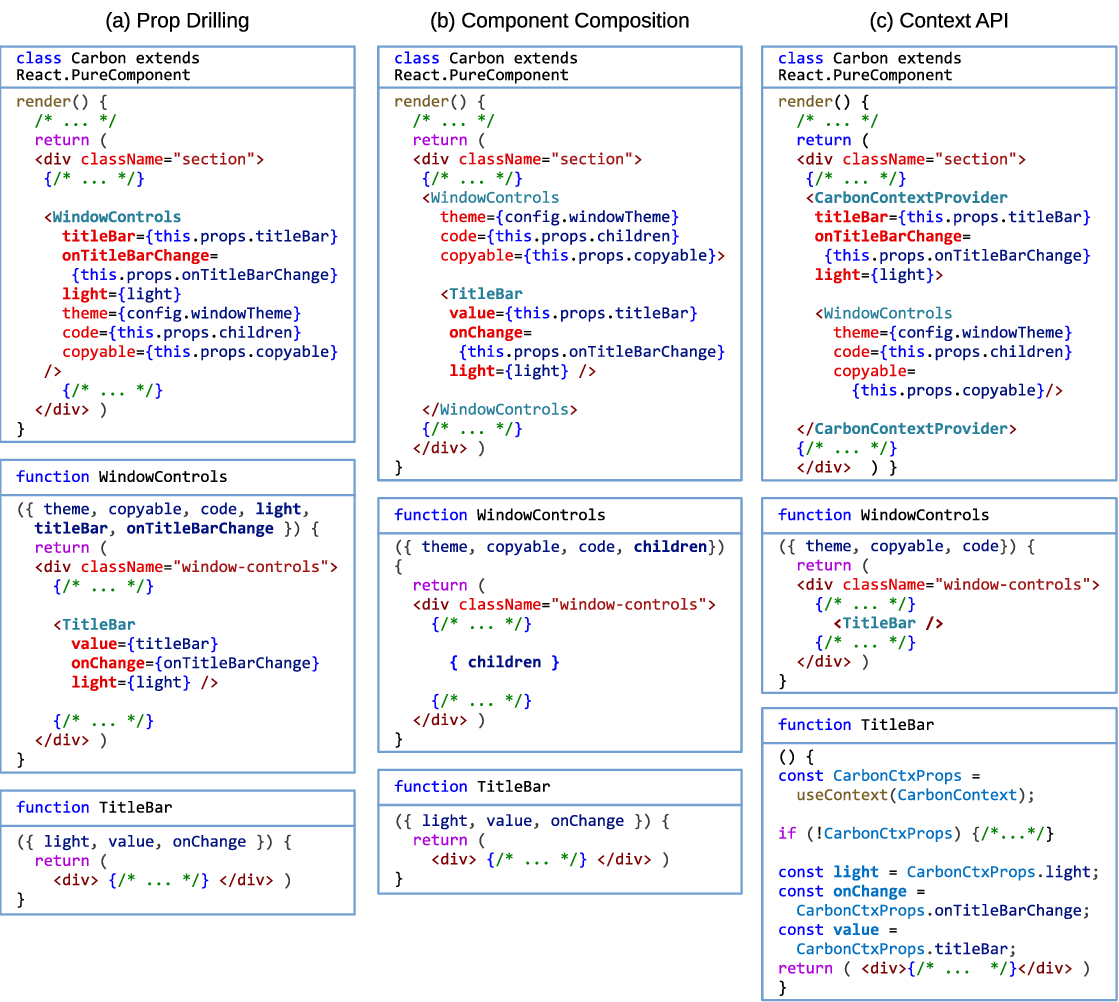}
    \captionsetup{justification=centering}
    \caption{Alternative designs based on Component Composition and the Context API.}
    \label{fig:alt-designs}
\end{figure*}

The official \React{} documentation discourages the use of \PropDrilling{}
for passing data deep in the component hierarchy~\cite{reactDevCtx},
due to verbosity and inconvenience in the declaration of components.
Moreover, a recent empirical study classifies \PropDrilling{} to code smells,
based on findings from a grey literature review,
validated through interviews with professional \React{} developers~\cite{Ferreira23}.
The rationale behind this characterization is twofold:
(a) difficulty in tracing the origin and uses of the forwarded property values,
especially, in large component hierarchies 
where components reside in different files,
(b) tight coupling among participant components,
since the forwarding of additional properties along the same path
requires changes to multiple components.
For instance, adding an extra property to the compound prop drilling from \Carbon{} to \TitleBar{}
(Figure~\ref{fig:prop-dril-example}) requires modification of three components.
The mitigation of \PropDrilling{} effects
can be attained with appropriate use of the \emph{Context API} and \emph{Component Composition};
native \React{} capabilities that facilitate the implementation
of adaptable and reusable components~\cite{reactDevCtx}.
Figure~\ref{fig:alt-designs} presents relevant code snippets from the components
participating in the compound prop drilling from \Carbon{} to \TitleBar{}
along with alternative designs based on component composition and the Context API.

\emph{Component Composition} facilitates component extensibility
by allowing child components to be nested within a component's JSX markup, 
thereby enabling the declarative construction of complex UI fragments.
A key enabler for component composition is the \children{} property
which represents the nested content passed to a component.
Components that expose this property can be extended by placing 
appropriate JSX markup between their opening and closing tags.
Figure~\ref{fig:alt-designs}(b) illustrates the employment of component composition
for the elimination of the compound prop drilling from \Carbon{} to \TitleBar{}.
Its target are the \lightProp{}, \valueProp{} and \onChange{} properties of \TitleBar{},
initially propagated from \Carbon{} through \WindowControls{}
(Figure~\ref{fig:alt-designs}(a)).
After refactoring, the \TitleBar{} markup 
is pulled up from \WindowControls{} to \Carbon{}
allowing for its direct configuration through prop passing.
The forwarded properties in \WindowControls{} are replaced by the \children{} property
and the \TitleBar{} markup by the content of \children{}.
Thus, \WindowControls{} ceases to depend on \TitleBar{}
and can be extended with other components having a similar role.

Despite being a straightforward approach,
recommended at first priority 
by the official \React{} documentation~\cite{reactDevCtx},
tackling a prop drilling with component composition
may require extensive refactoring of partipant components.
Actually, all components in the prop drilling path need to become \emph{composable},
in order to be appropriately nested and configured with prop passings
at the level of the source component.
By composable we mean extensible through nesting 
and without static dependencies among each other.
Thus, the component hierarchy along the prop drilling path is flattened,
target and intermediate components are decoupled 
and only the source component maintains dependencies to them.
The preconditions that allow the automation of this refactoring,
as well as the actual refactoring procedure
are studied in the next section.

\begin{figure}[!htb] 
    \centering
    \includegraphics[scale=0.8]{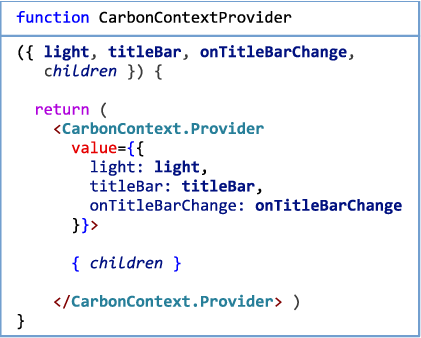}
    \captionsetup{justification=centering}
    \caption{Declaration of the CarbonContextProvider component.}
    \label{fig:ctx-provider}
\end{figure}

The elimination of \PropDrilling{} through the \emph{Context API}
has broader applicability.
It is based on the creation of a \emph{context} that makes a value,
either primitive or object,
available to a selected subtree of the render tree.
A context is created through the \createContext{} hook
and acts like a restricted variable scope.
Its reach is declared through a \ContextProvider{} component
that wraps the root of the subtree in JSX.
Components at any level of the subtree 
gain direct access to the context value,
eliminating, thus, the need for prop drilling.
Figure~\ref{fig:alt-designs}(c) illustrates 
the elimination of the compound prop drilling from \Carbon{} to \TitleBar{}
through the \emph{Context API}.
The context provider tag is applied in the JSX of the source component (\Carbon{})
and encloses \WindowControls{}, the target of the initial prop passing.
Notice, however, that \CarbonCtxProviderWrap{}
is a wrapper for the actual provider
that adapts its API to receive three properties
(\lightProp{}, \titleBar{}, \updateTitleBar{})
instead of a single \valueProp{} property (Figure~\ref{fig:ctx-provider}).
The \CarbonCtxProvider{} makes the context scope and its associated value
accessible by all components in the prop drilling path.
Thus, all properties involved in the prop drilling,
as well as their respective prop passings,
can be removed across the prop drilling path.
As presented in Figure~\ref{fig:alt-designs}(c),
the three forwarded properties are removed from the \WindowControls{} declaration
and are no longer provided to the \TitleBar{} tag.
Respective properties in the \TitleBar{} declaration,
initially serving as prop drilling target properties,
are, also, removed 
and \TitleBar{} receives the propagated data from the \CarbonContext{}.

In general, the use of the Context API 
incurs minimal changes to components in the prop drilling path,
except for the target component.
Especially for pure prop drillings,
the refactoring involves just the removal of forwarded properties
from component declarations and usages.
Changes to the target component focus on gaining access to the propagated data
through consumption of the Context API.
Figure~\ref{fig:alt-designs}(c) presents the refactored version of \TitleBar{}
that gains access to the \emph{value} provided to \CarbonCtxProvider{}
through the \useContext{} hook.
The hook receives an appropriate context object as argument (e.g., \CarbonContext{})
and, its successful outcome denotes that the component
is rendered as a descendant of the respective context provider. 
Removed properties from \TitleBar{}
are replaced by constants with the same name,
initialized with the respective properties
of the context value.
 
\section{Elimination of Prop Drillings through refactoring} 
\label{sect:refactoring}

In this section we introduce a method
for identifying \PropDrilling{} instances in a \React{} codebase
and eliminating them through automated refactoring.
The method employs static analysis
to recover a graph representation of the \React{} codebase
and introduces a graph traversal algorithm to identify prop drillings.
Moreover, we introduce a set of refactoring preconditions
that determine whether a \PropDrilling{} instance is refactorable
and which refactoring strategies are applicable -- Component Composition or Context API.
Finally, we specify source code transformation procedures
for eliminating a \PropDrilling{} instance through either refactoring strategy.

\subsection{Refactoring candidate identification procedure} 
\label{refactoringCandidateIdentificationProcedure}

The first stage in the automation of a refactoring 
is the identification of refactoring candidates,
i.e. code fragments that exhibit an undesired property or code smell
and, thus, warrant further evaluation for refactoring.
The proposed method treats 
all \PropDrilling{} instances as refactoring candidates.
Their identification requires the recovery
a graph representation of the \React{} codebase -- 
the Component Hierarchy Graph (CHG) -- 
which captures parent-child dependencies among components 
and the flow of data due to prop passings and prop forwardings.

\subsubsection{Component Hierarchy Graph construction}

\begin{figure*}[htb]
\begin{minipage}[t]{.45\textwidth}
  \begin{lstlisting}[style=styleA,
      captionpos=t,title={(a) Function Declaration}]
function Avatar({ person, size }) {
  return (
    <img className="avatar"
      src={getImageUrl(person)}
      alt={person.name}
      width={size}
      height={size} /> )
}
  \end{lstlisting}
  \end{minipage}
  \hfill
\begin{minipage}[t]{.5\textwidth}
  \begin{lstlisting}[style=styleA,
      captionpos=t,title={(b) Class Declaration}]
class Avatar extends PureComponent {
  render() {
    const { person, size } = this.props;
    return (
      <img className="avatar" ... /> )
  } 
}
  \end{lstlisting}
  \end{minipage}
  \hfill
\begin{minipage}[t]{.45\textwidth}
  \begin{lstlisting}[style=styleA,
      captionpos=t,title={(c) Variable declaration/Arrow function}]
const Avatar = ({ person, size }) => {
  return (
    <img className="avatar" ... /> )
}
  \end{lstlisting}
  \end{minipage}
  \hfill
\begin{minipage}[t]{.5\textwidth}
  \begin{lstlisting}[style=styleA,
      captionpos=t,title={(d) Variable declaration/Function expression}]
const Avatar = function({ person, size }){
  return (
    <img className="avatar" ... /> ) 
}
  \end{lstlisting}
  \end{minipage}
    \caption{Basic syntactic variants for \React{} component declaration in JS.}
    \label{fig:component-decl}
\end{figure*}

We construct the Component Hierarchy Graph (CHG) representation of the \React{} codebase
on the basis of components and their parent-child dependencies.
The graph is, further, enriched
with information regarding prop passings and prop forwardings
along component dependencies.
CHG construction involves static analysis of all production JS/TS files.
Since the dynamic features of JS/TS hinder the precise identification 
of components and their aliases,
the graph represents an approximation of the actual codebase.

\React{} components are identified 
by analyzing the Abstract Syntax Tree (AST) representation of each source file.
Despite the wide range of syntactic options for component declaration,
we assume that developers implement components, primarily,
through recommended practices of the official \React{} documentation.
Figures~\ref{fig:component-decl}(a) and~\ref{fig:component-decl}(b)
present the basic syntactic options for component declaration,
i.e., declaration as a JS function or an ES6 class, respectively.
Our method, also, supports identification of components
declared as arrow functions or function expressions
(Figures~\ref{fig:component-decl}(c),~\ref{fig:component-decl}(d)).

\begin{algorithm}[h!]
\caption{Component Identification Procedure}
\footnotesize
\label{alg:component-detection}
\begin{algorithmic}[1]
\Procedure{extractComponent}{ASTNode n, ASTNode t, Path p}

\State $supportedTypes \gets \{FunctionDeclaration,$
\Statex \hspace{5em} $ClassDeclaration,VariableDeclarator\}$

\If{$n.type \notin supportedTypes$}
    \State \Return null
\EndIf

\LComment{Ensure capitalized component names}
\State $name \gets n.id.name$
\If{$!\Call{isCapitalized}{name}$}
    \State \Return null
\EndIf

\LComment{Ensure variable declarations are initialized with function objects}
\State $allowedExpressions \gets \{FunctionExpression,$
\Statex \hspace{5em} $ArrowFunctionExpression\}$

\State \textbf{if} $n.type = VariableDeclarator\ \textbf{and}$
\Statex \hspace{2em} $n.init.type \notin allowedExpressions$\ \textbf{then}
\State \hspace{1em} \Return null

\LComment{Evaluate function/class declarations}
\State $baseClasses \gets \{Component, PureComponent\}$
\If{$n.type = ClassDeclaration$}
    \If{$n.superClass.name \notin baseClasses$}
        \State \Return null
    \EndIf
\ElsIf {$!\Call{returnsJSX}{n}$}
        \State \Return null
\EndIf

\LComment{Extract component details}
\State $fqn \gets \Call{extractFQN}{n, t, p}$
\State $props \gets \Call{extractProps}{n}$
\State $uses \gets \Call{extractComponentUses}{n, t}$

\State \Return $(n, name, fqn, props, uses)$

\EndProcedure
\end{algorithmic}
\end{algorithm}

Let $t$ be the AST representation of a JS/TS source file
conforming to the structure generated by widely adopted parsers 
(e.g. flow, acorn, babel).
The component identification algorithm traverses the tree structure $t$
and applies the \extractComponent{} procedure (Algorithm~\ref{alg:component-detection})
to appropriate types of AST nodes.
The procedure's input arguments comprise the currently analyzed AST node $n$,
the AST representation $t$, and
the source file's relative path $p$ within the codebase.
At first, the algorithm ensures that the AST type of node $n$
matches one of the syntactic variants of Figure~\ref{fig:component-decl},
i.e., \FunctionDeclaration{}, \ClassDeclaration{} or \VariableDeclarator{} (lines 2--4).
Then, it checks that $n$ has a capitalized name (lines 5--8),
a basic requirement for $n$ to be considered as a component
according to the \React{} specification.
In case that $n$ corresponds to a variable declaration,
the algorithm ensures that it is initialized with a function creation expression
i.e., an arrow function or a function expression (lines 9--12).
Up to this point, $n$ represents either a class or a function declaration.
In the former case, $n$ corresponds to a component declaration
if its parent class is one of the base classes of the \React{} library,
i.e., \Component{}/\PureComponent{} (lines 14--17).
Finally, a function declaration is identified as component,
if its return type is a JSX expression (lines 18--19).

The successful classification of an AST node $n$
as a \React{} component declaration,
is followed by further analysis of its AST subtree.
The analysis focuses on the extraction of component features
that are essential for the next steps
of the refactoring candidates' identification.
Algorithm~\ref{alg:component-detection} 
declares (lines 21--23) the extraction of each feature type
as a distinct analysis step,
represented by an appropriate function call.
For brevity reasons we omit their algorithmic descriptions.
Let $c$ be an identified component,
its extracted features comprise:

\begin{itemize}[leftmargin=*]

\item \emph{fqn}, a unique Fully Qualified Name (FQN) for the component, 
required for resolution of component dependencies.
It is formed through concatenation of the path $p$ 
with the component $name$, i.e. $p:name$. 
Function \extractFQN{} (line 21) analyzes the AST $t$
to infer the component FQN.

\item \emph{props}, the set of component properties,
declared either directly in its function declaration,
using the object destructuring syntax,
or inferred through analysis of the component implementation.
In the latter case,
we infer properties from references to the \code{this.props} object.
This analysis is performed by the \extractProps{} function (line 22).

\item \emph{uses}, the set of component uses
corresponding to JSX tags in the returned markup of component $c$.
We exclude uses of components 
that are either part of the \React{} ecosystem,
or are imported from external libraries.
Each component use $cu_i$ is a tuple of the form 
$(name_i, fqn_i, attr_i)$,
where (a) $name_i$ is the component name as extracted from the JSX tag,
(b) $fqn_i$ is the FQN, identified through analyzing all import statements 
and, (c) $attr_i$ is the set of property assignments 
retrieved from the JSX attributes of the tag $cu_i$.
Each property assignment is a pair $(prop, value)$,
where $prop$ is the JSX attribute name 
and $value$ is the assigned value.
The analysis of component uses 
is performed by function \extractUses{} (line 23).

\end{itemize}

A component identified by the \extractComponent{} procedure,
along with its features,
is returned in the form of a tuple (line 24).
The application of the procedure 
to the ASTs of all JS/TS files of the codebase
results to the set of components $C$.
We will, henceforth, refer to the features 
of a component $c \in C$ through dot notation,
e.g. $c.n$ for the AST node, $c.name$ for its name, etc.

The \emph{Component Hierarchy Graph} is a directed graph $CHG=(C, E)$,
where the nodeset $C$ corresponds to the components of the codebase
and the set of edges $E$ represent the parent-child dependencies among them.
A parent-child dependency denotes that
the parent component renders the child component,
or equivalently, the child component appears
in the \emph{uses} of the parent component.
In the rest of this section we will use the notation $E_{c}$
to refer to a subset of $E$
that includes all edges originating from the component $c$.

\begin{algorithm}[h!]
\caption{Edge Resolution Procedure}
\footnotesize
\label{alg:edge-resolution}
\begin{algorithmic}[1]
\Procedure{resolveEdges}{Set C}
\State $E \gets \emptyset$
\ForAll{$c_i \in C$}
    \ForAll{$u \in c.uses$}
        \LComment{Resolve the component that corresponds to the FQN}
        \State $c_u \gets f(u.fqn)$
        \State $(PP_{i, u}, PF_{i, u}) \gets \Call{resolvePropForwardings}{c_i, c_u, u.attr}$
        \LComment{Create a new edge and append it to the set of edges}
        \State $e \gets (c_i, c_u, PP_{i,u}, PF_{i, u})$
        \State $E \gets E \bigcup \{e\}$
    \EndFor
\EndFor
\State \Return $E$
\EndProcedure
\end{algorithmic}
\end{algorithm}

The construction of the set $E$ 
is outlined in Algorithm~\ref{alg:edge-resolution}.
Specifically, the \resolveEdges{} procedure processes each component $c \in C$
and introduces a new edge for each component use $u$ in $c.uses$.
The function $f$ maps the FQN $u.fqn$ of the referenced component 
to a unique node $cu \in C$ (line 6).
Prop forwardings from $c$ to $cu$ are identified 
through invocation of the \resolvePropForwardings{} function
that receives as arguments the component representations,
as well as the property assignments on component use $u$ (line 7).
The details of prop forwarding resolution will be explained hereafter.
Finally, a new edge representation is created and added to the set of edges (lines 8--9).
Note that each edge is represented as a tuple $(c_a, c_b, PP_{a, b}, PF_{a,b})$,
where $c_a$, $c_b$ are the parent and child components,
while $PP_{a, b}$, $PF_{a,b}$ represent 
the sets of prop passings and prop forwardings from $c_a$ to $c_b$, respectively.

\begin{figure}[htb] 
    \centering
    \includegraphics[scale=0.8]{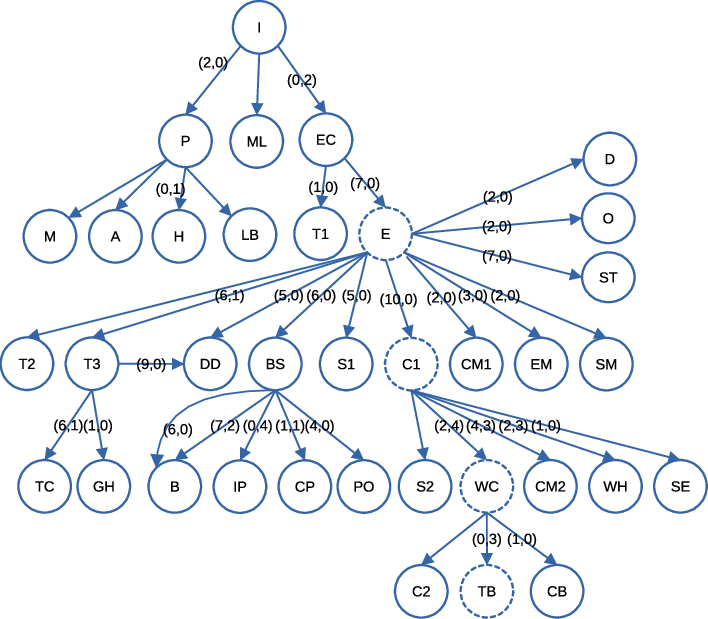}
    \captionsetup{justification=centering}
    \caption{Fragment of the CHG for the \Carbon{} project.}
    \label{fig:component-hierarchy}
\end{figure}

Figure~\ref{fig:component-hierarchy} presents 
a fragment of the CHG for the \carbon{} project.
For illustration purposes,
each node is tagged with the first letter of the component name 
(e.g. \code{I} for \code{Index})
or the first letters of the parts of a compound name
(e.g. \code{WC} for \WindowControls{}).
For component names starting with the same letter,
we combine the starting letter with an appropriate digit.
Nodes highlighted with a dashed outline
correspond to the components of Figure~\ref{fig:prop-dril-example}.
Each edge is tagged 
with the cardinalities of the sets $PP$ and $PF$ inside parentheses.
Edges without tag denote that 
the cardinality of boths sets is zero, i.e. \code{(0,0)}.
For instance, 
edge \code{(2,4)} among components \code{C1} and \code{WC} denotes that
\Carbon{} sets values to 6 properties of the \WindowControls{} tag in the returned JSX markup;
two of them are plain prop passings and 
the rest are forwardings of \Carbon{} properties.

\subsubsection{Resolution of Prop Forwardings}

The resolution of prop passings and prop forwardings from a component $c_i$
to a specific use of component $c_u$ in the returned JSX of $c_i$
is performed by procedure \resolvePropForwardings{}.
The procedure receives as arguments the representations of components $c_i$, $c_u$,
as well as the set of property assignments $attr$ provided to the respective JSX tag.
Property assignments are processed to construct
the sets of prop passings $PP_{i, u}$ and prop forwardings $PF_{i, u}$
from component $c_i$ to a use of component $c_u$.
The procedure analyzes each property assignment $(p_j, value),\ p_j \in P_u$
contained in $attr$ through the following rules:

\begin{itemize}
    \item if $value$ is a constant value 
    or a complex curly braces expression (e.g. \code{\{expr\}}), 
    then the property assignment is classified as a plain prop passing
    $PP_{i,u,j} = (c_i, c_u, p_j)$ that is added to the set $PP_{i, u}$,
    
    \item if $value$ is a simple curly braces expression
    of the form \code{\{var_name\}}, 
    where \code{var_name} is a simple name,
    then we identify the following cases for \code{var_name}:

    \begin{itemize}
        \item \code{var_name} corresponds to a property $p_k \in P_i$
        that is not referenced in any other location in the implementation of $c_i$.
        In this case a new \emph{pure} prop forwarding $\overline{PF}_{i,u,k,j} = (c_i, c_u, p_k, p_j)$
        is added to $PF_{i, u}$.
        \item \code{var_name} corresponds to a property $p_k \in P_i$
        that is referenced without being modified
        in one or more locations in the implementation of $c_i$.
        In this case a new prop forwarding $PF_{i,u,k,j} = (c_i, c_u, p_k, p_j)$
        is added to the set $PF_{i, u}$.
        \item in any other case, 
        a new prop passing $PP_{i,u,j} = (c_i, c_u, p_j)$ 
        is added to the set $PP_{i, u}$.
    \end{itemize}
    
\end{itemize}

The sets $PP_{i, u}$, $PF_{i, u}$ are, finally, 
returned by the procedure in a tuple form.

\subsubsection{Identification of \PropDrilling{} instances}

The identification of \PropDrilling{} instances is based on CHG traversal.
Graph traversal follows sequences of compatible prop forwardings
along successive edges to construct \PropDrilling{} instances.
The search is initiated from root components,
i.e., components serving as entry points to the rendered component hierarchy.
This ensures that suggested refactorings involve components 
reachable from the application roots during rendering.
Root components can be inferred from common code-layout conventions 
(e.g. component \code{App} defined in an \code{index.js} file)
or explicitly provided as user input.
For a more exhaustive search,
components without incoming edges in the CHG
can also be considered as root components.

\begin{algorithm}[htb!]
\caption{\PropDrilling{} identification procedure}
\footnotesize
\label{alg:prop-drilling}

\begin{algorithmic}[1]
\Procedure{identifyPropDrillings}{Set R}
\State $PD \gets \emptyset$
\State $pfPaths \gets \emptyset$
\LComment{Find prop forwarding paths }
\ForAll{$r \in R$}
    \State $paths \gets findPFPaths(r)$
    \State $pfPaths \gets pfPaths \bigcup paths$
\EndFor
\LComment{Search for PDs within prop forwarding paths}
\ForAll{$p \in pfPaths$}
    \State $pathPD \gets searchPropDrillings(p)$
    \State $PD \gets PD \bigcup pathPD$
\EndFor

\State \Return $PD$

\EndProcedure
\end{algorithmic}
\end{algorithm}

Algorithm~\ref{alg:prop-drilling} outlines the \identifyPropDrillings{} procedure.
The procedure receives as argument the set $R$ of application root components
and returns a set of \PropDrilling{} instances.
At first, the algorithm constructs a set of \emph{Prop Forwarding paths} (PF-paths)
through depth-first traversal of the CHG.
A PF-path is a directed path that starts either from 
(a) an outgoing edge of a root component or
(b) an edge with no prop forwardings, 
i.e. an edge tagged with \code{(*,0)} in Figure~\ref{fig:component-hierarchy}.
Since a prop drilling has at least three participant components,
the length of a PF-path $p$ should be at least 2, i.e. $|p| \geq 2$.
A PF-path should only contain edges with prop forwardings,
except for its first edge.
It terminates to a CHG node that either has no outgoing edges,
or its outgoing edges contain no prop forwardings (e.g. path \code{I-EC}).
For instance, path \code{E-C1-WC-TB} in Figure~\ref{fig:component-hierarchy},
starts from edge \code{E-C1} with no prop forwardings and 
terminates to edge \code{WC-TB} containing 3 prop forwardings to component \TitleBar{}.
All prop drillings contained in the example of Figure~\ref{fig:prop-dril-example}
are identified along this path.

\begin{algorithm}[htb!]
\caption{\PropDrilling{} search procedure}
\footnotesize
\label{alg:pd-search}
\begin{algorithmic}[1]
\Procedure{searchPropDrillings}{Path pfPath}

\State $pdSet \gets \emptyset$
\State $n \gets |pfPath|$
\LComment{Iterate till the penultimate edge of the path}
\ForAll{$e_i \in pfPath: i \in [1,n-1]$}
    \LComment{Unpack the contents of the edge to variables}
    \State $(c_a, c_b, PP_{a,b}, PF_{a,b}) \gets e_i$
    \State $PP_i \gets PP_{a,b}$
    \If{$\Call{isRoot}{c_a}$}
        \State $PP_i \gets PP_i \bigcup PF_{a,b}$
    \EndIf
    \LComment{Search for a prop drilling from each prop passing}
    \ForAll{$pp \in PP_i$}
        \State $(\_, \_, p) \gets pp$
        \LComment{Initialize the prop drilling with a prop passing}
        \State $pd \gets (pp)$
        \ForAll{$e_j \in pfPath: j \in [i+1,n]$}
            \State $(\_, \_, \_, PF_j) \gets e_j$
            \State $pf \gets \Call{getMatchingPF}{p, PF_j}$
            \If{$pf = null$}
                \LComment{Prop forwarding sequence terminated}
                \If{$|pd| \geq 2$}
                    \State $pdSet \gets pdSet \bigcup \{pd\}$
                \EndIf
                \State \textbf{break}
            \EndIf
            
            \State $pd \gets pd \cdot pf$
            \State $(\_, \_, \_, p_t) \gets pf$
            \State $p \gets p_t$
        \EndFor
     \EndFor
\EndFor
\State \Return $pdSet$
\EndProcedure
\end{algorithmic}
\end{algorithm}

The identification of prop drillings along a given PF-path
is performed by procedure \searchPropDrillings{},
which is outlined in Algorithm~\ref{alg:pd-search}.
The procedure processes sequentially the edges of the PF-path
and analyzes their sets of prop passings (lines 5--26).
Specifically, starting from each prop passing,
the algorithm follows matching prop forwardings
across the edges of the PF-path to construct a prop drilling (lines 13--26).
The last edge of each PF-path is not processed,
since a prop drilling should contain at least one prop forwarding. 
Each edge $e_i$ of the CHG has a tuple representation 
and its processing requires unpacking its elements to appropriate variables (line 7).
Notice that the tuple ``destructuring'' notation used in line 7
will be, also, applied in other parts of the algorithm (e.g. line 13).
The set of prop passings of an edge $e_i$ 
is assigned to variable $PP_i$ (line 8).
The set $PP_i$ is, further, expanded 
with the prop forwardings of the current edge,
in cases that the edge originates from a root component (lines 8--10).
The reason is that root components may forward props
provided outside the component scope,
e.g. by code that renders the root component
in the host page of the \React{} application.

The actual processing of each prop passing $pp \in PP_i$
is specified in lines 13--26 of the algorightm.
At first, the name of the prop passing's target property 
is extracted to a variable $p$ (line 13).
Notice the use of the \code{'\_'} placeholder in the tuple destructuring notation
to refer to elements not required for further processing.
Then, a new prop drilling is initiated as a single element sequence,
containing the prop passing $pp$ (line 15).
The prop drilling is constructed
by following the edges of the PF-path,
starting from edge $e_{i+1}$ (lines 16--26). 
Initially, the property name $p$ is matched 
against the set of prop forwardings $PF_{i+1}$
of the next edge $e_{i+1}$ (lines 17--18).
Matching is performed by function \getMatchingPF{}
and involves finding a prop forwarding $pf \in PF_{i+1}$
that has $p$ as its source property.
The function receives as arguments the property name $p$
and the set of prop forwardings $PF_{i+1}$ (line 18).
In case of successful matching the function returns a prop forwarding $pf$
that is added to the sequence $pd$ (line 24).
The algorithm, then, proceeds with the next edge in the PF-path
having as property to be matched,
the target property of the identified prop forwarding $pf$ (lines 25--26).
In case of a failed matching,
the algorithm completes the processing of $pp$
and evaluates the sequence $pd$.
If the sequence $pd$ has at least 2 prop passings (line 21),
then it is considered as a prop drilling
and is added to the set $pdSet$ of identified prop drillings (line 22).

\subsection{Source Code Transformation Procedure} 
\label{sourceCodeTransformationProcedure}

In this section we specify source code transformation procedures
for eliminating a \PropDrilling{} instance
through \emph{Component Composition} or the \emph{Context API}.
We focus on the general case of a compound prop drilling $CPD_{s,t,J_t}$
that forwards a set of properties $J_t\subseteq P_t$ to the target component $C_t$
through $N=|J_t|$ prop drillings sharing the same path
$(C_s, C_0, \ldots, C_{n-1}, C_t)$.
The method can be adapted for individual prop drillings by setting $N=1$.
For brevity reasons,
we will use the notation $\pure{J_i} \subseteq J_i$ 
to refer to the properties of component $C_i$
that are forwarded through \emph{pure} prop forwardings.
Moreover, $\nonpure{J_i} \equiv J_i \setminus \pure{J_i}$.
For the target component $C_t$, 
we assume that $\pure{J_t} = J_t$ and $\nonpure{J_t} = \emptyset$.

Our method first evaluates 
the applicability of the \emph{Component Composition} strategy
against a set of refactoring preconditions.
These preconditions ensure that
the refactoring does not introduce syntax errors
or alter the external behaviour of the program.
If any precondition is violated,
the applicability of the \emph{Context API} strategy is evaluated
against a separate set of preconditions.
If any of these preconditions is violated
the \PropDrilling{} instance is rejected 
as a candidate for automated refactoring.

\subsubsection{Refactoring Preconditions}

The refactoring preconditions that ensure safe elimination
of a compound prop drilling $CPD_{s,t,J_t}$ 
through Component Composition are outlined below:

\medskip

\begin{enumerate}
[labelindent=0cm,
				style=multiline,
				leftmargin=*,
				label=\emph{{RP\arabic*}}]

\item The use of the target component $C_t$ within the JSX of its parent component $C_{n-1}$
      should receive prop passings exclusively for the forwarded properties $J_t$.
      The precondition ensures that the JSX tag for $C_t$
      can be safely removed from the returned JSX of $C_{n-1}$
      and replaced by an appropriate placeholder JSX expression
      during refactoring.
      The precondition applies recursively 
      to all intermediate components $C_0, \ldots, C_{n-1}$.

\item Each intermediate component $C_i, i \in 0,\ldots,n-1$
      should have a single incoming edge in the CHG 
      corresponding to its parent component $C_{i-1}$.
      The presence of multiple incoming edges denotes 
      that $C_i$ is used in the returned JSX of additional components.
      Thus, removal of the forwarded properties from the declaration of $C_i$
      may introduce errors, due to potential prop passings to these properties
      from other component uses.       

\item Given any component $C_i \in \{C_1, \ldots, C_t\}$,
      $C_i$ must be used in its parent component $C_{i-1}$
      exclusively as a tag in the returned JSX by $C_{i-1}$.
      Other forms of use, e.g. \code{Component = wrapper(Component)}, 
      may alter the behaviour of $C_i$ 
      once it is moved to the scope of the source component $C_s$.
      Notice that this precondition does not apply to component $C_0$
      which remains in the scope of $C_s$ after refactoring.

\item Given any component $C_i \in \{C_1, \ldots, C_t\}$,
      declared in a separate file from $C_s$,
      its import in the $C_s$ source file should not 
      introduce conflicts with other imported/local identifiers.
      Similar to RP3, 
      the precondition does apply to $C_0$ 
      which is already present in the $C_s$ source file.

\item Given any intermediate component $C_i \in \{C_1, \ldots, C_{n-1}\}$,
      it should not declare and use in its implementation 
      the predefined \code{children} property.
      Since expression \code{\{children\}} will be used as placeholder
      for component $C_{i+1}$ in the returned JSX of $C_i$,
      its presence in other locations 
      will alter the presentation of $C_i$ after refactoring.
    
\end{enumerate}

\medskip

Refactoring through the Context API 
requires a less extensive set of preconditions,
which are summarized below:

\begin{enumerate}
[labelindent=0cm,
				style=multiline,
				leftmargin=*,
				label=\emph{{RP\arabic*}}]
\setcounter{enumi}{5}

\item Extend RP2 and apply its restrictions to the target component $C_t$, as well.
      The reason is that the current refactoring strategy
      involves removing all forwarded properties contained in $J_t$
      from the declaration of the target component $C_t$.
      Thus, any prop passings to these properties 
      from other components would introduce runtime errors.
          
\item Import of the context provider wrapper component
      in the source file of $C_s$ should not 
      introduce conflicts with other imported/local identifiers.

\end{enumerate}

\subsubsection{Refactoring to Component Composition}

The source code transformation procedure
for elimination of a compound prop drilling $CPD_{s,t,J}$ through \emph{Component Composition}
applies to components $C_s, C_0, \ldots, C_{n-1}$
and primarily modifies property declarations and their returned JSX markup.
The target component $C_t$ is not affected by the refactoring.
The transformation procedure comprises the following steps:

\begin{figure*}[htb] 
    \centering
    \includegraphics[scale=\listingScale]{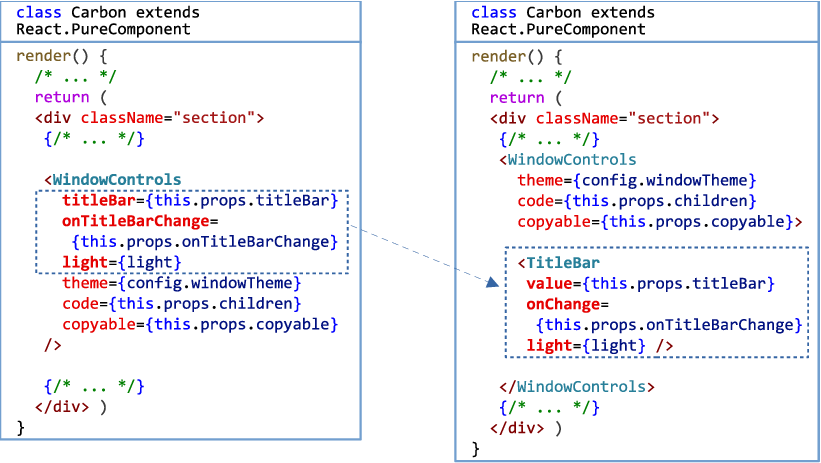}
    \captionsetup{justification=centering}
    \caption{Carbon component before and after refactoring with Component Composition.}
    \label{fig:ref-cc-carbon}
\end{figure*}

\paragraph{Step 1} Transform the returned JSX of the source component $C_s$.

\begin{enumerate}[label=(\alph*)]
  \item Introduce component uses (tags) for components $C_1 \ldots C_{n-1}, C_t$.
  For each $C_i \in \{C_1, \ldots, C_{n-1}, C_t\}$ 
  alter the returned JSX of $C_s$ as follows:
  \begin{itemize}
    \item Nest a tag for $C_i$ within the tag of component $C_{i-1}$,
      e.g., the tag for component $C_1$ will be nested in the tag for $C_0$,
      resulting to the markup \code{<C0><C1/></C0>}.
    \item Introduce a prop passing in tag $C_i$
      for each property $p \in \nonpure{J_i}$,
      corresponding to a non pure prop forwarding.
      The assigned value is copied 
      from the respective prop passing to component $C_0$
      that initiated the prop drilling for property $p$.
  \end{itemize}
  
  \item Remove all prop passings corresponding to $\pure{J_0}$ 
    from the $C_0$ tag.
    Prop passings for the rest of the forwarded properties are preserved,
    since their values are used in other locations within $C_0$.
\end{enumerate}

To better illustrate the effects of Step 1, 
assume that all prop passings from component $C_s$ to $C_0$ are propagated to $C_t$
exclusively through pure prop forwardings.
Then, the tag \code{<C0 p1="v1" p2="v2" ... pN="vN"></C0>} in the returned JSX of $C_s$
is replaced with the following markup:

\begin{lstlisting}
  <C0>
    <C1>
      ... <CN-1>
              <Ct p1="v1" p2="v2" ... pN="vN"/>
      ... </CN-1>
    <C1>
  </C0>
\end{lstlisting}

Figure~\ref{fig:ref-cc-carbon} presents the application of Step 1
to the compound prop drilling from \Carbon{} to \TitleBar{}.
As part of Step 1(a), the tag of intermediate component \WindowControls{} is modified 
to nest the tag of the target component \TitleBar{}.
Moreover, all prop passings 
initially provided to \WindowControls{} tag 
are applied to \TitleBar{} with modified property names.
Finally, as part of Step 1(b) properties 
\titleBar{}, \onTitleBarChange{} and \lightProp{},
involved in the prop drilling, are removed from WindowControls{}.

\begin{figure*}[htb] 
    \centering
    \includegraphics[scale=\listingScale]{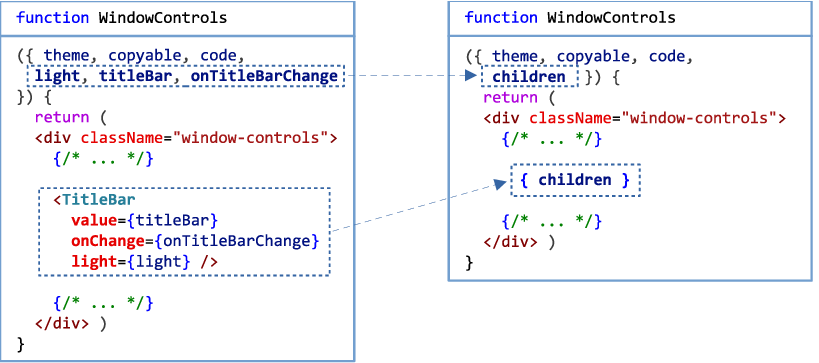}
    \captionsetup{justification=centering}
    \caption{WindowControls before and after refactoring with Component Composition.}
    \label{fig:ref-cc-wc}
\end{figure*}

\paragraph{Step 2} Transform each intermediate component $C_i$.
\begin{enumerate}[label=(\alph*)]
    \item Remove property declarations for all properties in $\pure{J_i}$.
    \item Introduce a property declaration 
          for the reserved \code{children} property
          to the component declaration.
    \item Replace the component $C_{i+1}$ tag (or $C_t$ for $i=n-1$)
    with the \code{\{children\}} JSX expression.
\end{enumerate}

Figure~\ref{fig:ref-cc-wc} illustrates 
the elimination of three prop forwardings 
from the refactored version of \WindowControls{},
as well as the removal of its dependency to the \TitleBar{} component.
The \WindowControls{} component is, now, 
transformed to a \emph{composable} component 
that can be configured through tag nesting
with any markup that implements content related to window controls.

\subsubsection{Refactoring to Context API}

The refactoring of $CPD_{s,t,J}$, through the Context API, 
applies to all participant components.
It, also, introduces an adapter component 
that adapts \ContextProvider{} 
to receive multiple custom properties 
in place of the \valueProp{} property.
The transformation procedure involves the following steps:

\begin{figure*}[htb] 
    \centering
    \includegraphics[scale=\listingScale]{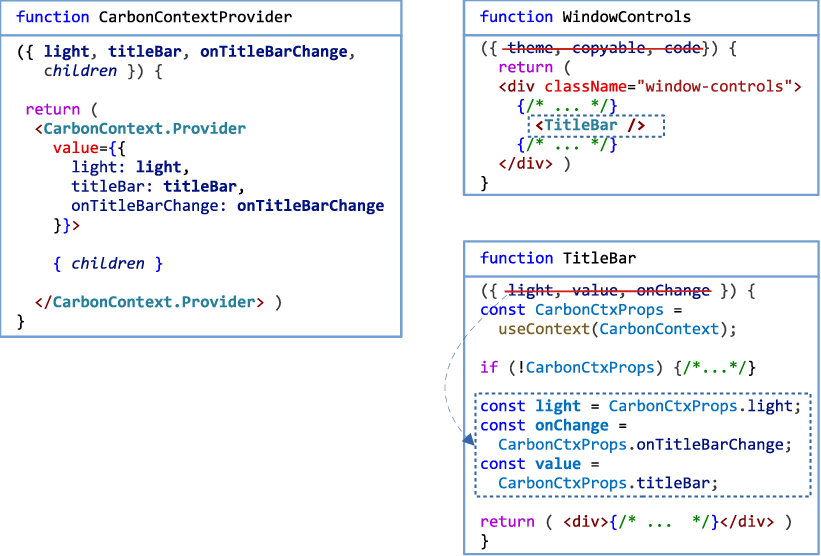}
    \captionsetup{justification=centering}
    \caption{Adapter component and refactored versions of WindowControls and TitleBar.}
    \label{fig:ref-ctx-wc-tb}
\end{figure*}

\paragraph{Step 1} Create and export the \emph{Context} object.
    
\begin{enumerate}[label=(\alph*)]
  \item Create the \code{<NameS>-context.jsx} file, 
  where \code{<NameS>} is the name of the source component $C_s$.
  \item Create the \code{<NameS>Context} object in this file, 
  through the \code{createContext()} hook and export it with the same name.
\end{enumerate}

\paragraph{Step 2} Create and export an adapter for the context provider component.

\begin{enumerate}[label=(\alph*)]

    \item Declare a functional component named \code{<NameS>ContextProvider}
    in file \code{<NameS>-context.jsx}.
    \item Introduce the forwarded properties $J_0$ 
    and the reserved \code{children} property in its declaration.
    \item Nest the \code{\{children\}} expression inside the \ContextProvider{} tag.
    Return the latter as the JSX specification of \code{<NameS>ContextProvider}.
    \item Assign to the \valueProp{} attribute of \ContextProvider{}
    an object literal with all properties in $J_0$.
    Initialize each property with a reference
    to the homonymous property \code{<NameS>ContextProvider}.  
\end{enumerate}

The left part of Figure~\ref{fig:ref-ctx-wc-tb} presents 
the \CarbonCtxProviderWrap{} adapter.
The \code{children} property allows for nesting within an adapter tag
subtree of components that require access to the context properties.

\paragraph{Step 3} Transform the returned JSX of the source component $C_s$.
\begin{enumerate}[label=(\alph*)]
    \item Nest the tag $C_0$ within the \code{<NameS>ContextProvider} tag.
    \item Move all property assignments, corresponding to properties in $J_0$,
    from the $C_0$ tag to the \code{<NameS>ContextProvider} tag.
\end{enumerate}

\begin{figure*}[htb] 
    \centering
    \includegraphics[scale=\listingScale]{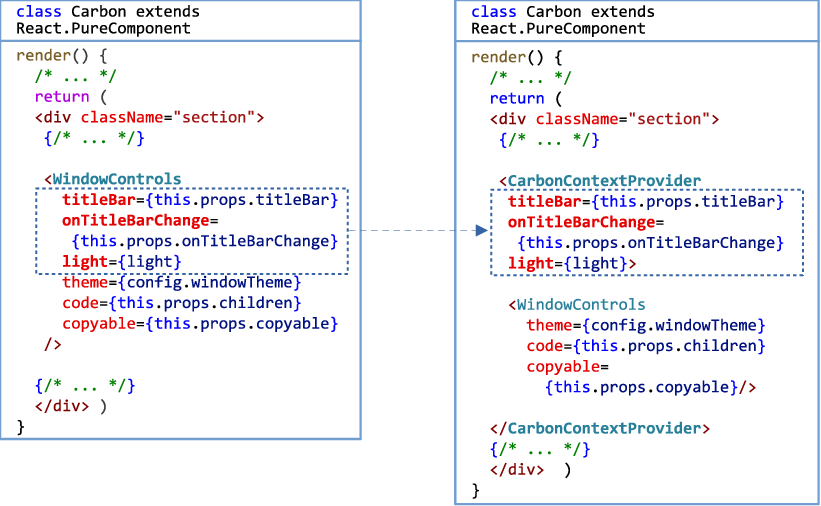}
    \captionsetup{justification=centering}
    \caption{Carbon component before and after refactoring through the Context API.}
    \label{fig:ref-ctx-carbon}
\end{figure*}

Figure~\ref{fig:ref-ctx-carbon} illustrates the \Carbon{} component
before and after refactoring with the Context API.
Notice that the prop passings initiating the compound prop drilling
are moved to the adapter component
and the respective property values become available
to all components of the \WindowControls{} subtree.

\paragraph{Step 4} Transform the implementation of the rest of
participant components.
For each $C_i \in (C_0, C_1, \ldots, C_t)$ do the following:
\begin{enumerate}[label=(\alph*)]
    \item Remove all forwarded properties from the component specification.
    \item Remove all prop forwardings to component $C_{i+1}$,
    unless $C_i = C_t$.
    \item In case that $\nonpure{J_i} \neq \emptyset$, 
    assign to the \code{<NameS>CtxProps} variable 
    the value of \code{<NameS>Context} through the \code{useContext()} hook.
    For each $p \in \nonpure{J_i}$ do:
    \begin{itemize}
      \item Declare a local variable with the same name 
      and initialize it with the respective property of the \code{<NameS>CtxProps} object.
    \end{itemize}
\end{enumerate}

The right part of Figure~\ref{fig:ref-ctx-wc-tb} presents 
the refactored versions of \WindowControls{} and \TitleBar{}.
The specification of both components is simplified
through removal of all forwarded properties.
In case that \WindowControls{} involved 
one or more non-pure prop forwardings, 
its implementation would include similar statements
with the first five statements of \TitleBar{}.

\section{Empirical evaluation} 
\label{sect:evaluation}

In this section we present an empirical study
of our method for automated identification
and elimination of prop drillings in \React{} codebases.
The \emph{goal} of this study,
according to the goal template of~\citet{Wohlin2000}, 
is to analyze the proposed method
for the \emph{purpose} of evaluation
\emph{with respect to} applicatility, effectiveness and practicality.
The study is conducted from the \emph{perspective}
of (a) a web developer seeking to understand 
the prevalence and complexity of the \PropDrilling{} code smell,
as well as its potential for elimination through automated refactoring;
and (b) an IDE tool developer
evaluating potential refactoring capabilities 
for integration into web development tools.
The \emph{context} of the study is
a set of open-source \React{} applications.

\subsection{Research Questions}

The experimental study aims to answer the following research questions:

\emph{RQ1. How common are Prop Drilling occurrences in real-world \React{} applications?} 
The research question aims to investigate the extent 
to which \PropDrilling{} is used 
for data propagation along the component hierarchy.
The results of this research question will provide insights
into the scope and applicability of the proposed refactoring.

\emph{RQ2. What is the range of Prop Drilling complexity?}
Our purpose is to explore the complexity of \PropDrilling{} instances
in terms of (a) depth, i.e., number of intermediate components,
and (b) breadth, i.e., number of properties 
that are simultaneously forwarded along the same path.
The results of this research question 
will highlight the effectiveness of the proposed method
in eliminating unnecessary data dependencies
and coupling among components.

\emph{RQ3. What ratio of Prop Drillings can be eliminated through fully automated refactoring?}
The research question aims to evaluate the potential 
for fully automated refactoring of identified \PropDrilling{} instances,
either through the Component Composition or the Context API strategy.
Results will support the practicality of the proposed method,
since its implementation in IDEs would enable refactoring automation 
in several cases of poor design,
freeing the developer from the tedious and error-prone task
of manual refactoring.

\subsection{Context Selection}

The context of the empirical study comprises a dataset of 15 \React{} projects,
openly available as Github repositories.
Each repository implements either a \React{} application or
a full-stack web application with a \React{} front end.
To construct the dataset,
we collected all projects included in the empirical studies
of ~\citet{Ferreira23},~\citet{Ferreira24} and ~\citet{Nunes2025}.
These studies are the most closely related to our work
and were published in top-tier journals.
From the 30 projects considered across the three studies,
we identified 19 distinct projects
and excluded four because they implement component libraries 
(e.g. \proj{ant-design}, \proj{material-ui}).
Component libraries primarily comprise
declarations of reusable, isolated components and,
thus, their internal structure differs substantially 
from that of a typical \React{} application.

Table~\ref{tbl:projects} presents implementation details
for all projects included in our dataset,
hereafter referred to as ``benchmark'' projects.
Columns 2 and 3 provide the implementation language(s)
and SLOC (Source Lines Of Code) size
for each project\footnote{Computed by CLOC tool, https://github.com/AlDanial/cloc}.
For SLOC size estimation,
we exclude configuration files and
source files implementing non front-end responsibilities.
Column 4 presents the number of GitHub stars for each project, 
while Column 5 provides, for reproducibility purposes,
the Commit Id of the analyzed snapshot of each codebase.
Finally, Column 6 includes the projects' Github repository names.

\begingroup
\setlength{\tabcolsep}{6pt} 
\renewcommand{\arraystretch}{1.3} 
\begin{table*}[htb]
\centering
\scriptsize
\begin{tabular}{l l r r r l}

\multicolumn{1}{c}{\theading{Project}} &
\multicolumn{1}{p{0.9cm}}{\theading{Language}} &
\multicolumn{1}{c}{\theading{SLOC}} &
\multicolumn{1}{c}{\theading{Stars}} &
\multicolumn{1}{c}{\theading{Commit Id}} &
\multicolumn{1}{c}{\theading{Github Repository}} \\ \hline

grafana         & TS    & 696744 	& 76.6K 	&  0d1a5b4 	& grafana/grafana        \\
metabase        & JS/TS & 477532 	& 49.1K 	&  7fbf305 	& metabase/metabase      \\
affine          & TS    & 231911 	&  72.5 	&  5c0e3b8 	& toeverything/AFFiNE    \\
superset        & JS/TS & 205589 	& 74.6K 	&  6a1c30e 	& apache/superset        \\
novu            & TS    & 170978 	&   40K 	&  cfbea2d 	& novuhq/novu            \\
lobe-chat       & TS    & 169644 	&  82.5 	&  a11925f 	& lobehub/lobehub        \\
plane           & TS    & 143712 	&  59.4 	&  e710f5b 	& makeplane/plane        \\
Rocket.Chat     & TS    & 142439 	& 46.1K 	&  7137ec8 	& RocketChat/Rocket.Chat \\
cal.com         & TS    &  89443 	&  48.4 	&  947b05b 	& calcom/cal.com         \\
mastodon        & JS/TS &  56721 	& 50.3K 	&  55a7b1e 	& mastodon/mastodon      \\
joplin          & JS/TS &  47341 	& 56.2K 	&  e75875c 	& laurent22/joplin       \\
redash          & JS/TS &  24530 	& 28.8K 	&  67a95e9 	& getredash/redash       \\
mitmproxy       & JS    &  16540 	& 44.9K 	&  d9d5878 	& mitmproxy/mitmproxy    \\
prometheus      & JS/TS &  13516 	& 65.9K 	&  9ec59ba 	& prometheus/prometheus  \\
carbon          & JS    &  11580 	& 36.1K 	&  0a07790 	& carbon-app/carbon      \\
\cmidrule{1-6}
\end{tabular}

\captionsetup{justification=centering}
\caption{\React{} projects used in the empirical evaluation.}
\label{tbl:projects}
\end{table*}
\endgroup

The proposed method is implemented as a Node.js command line tool,
\ReactRefactor{},
published in Github\footnote{https://github.com/Vginis/ReactRefactor}
through an open-source license.
Its implementation is based on \jscodeshift\footnote{https://github.com/facebook/jscodeshift}
for parsing \React{} source code to an Abstract Syntax Tree (AST) representation
and producing the refactored code through AST transformations.
The tool requires minimal user input, 
primarily the root directory of the \React{} front-end implementation of the analyzed codebase.
After the identification of refactoring candidates,
the tool provides a list preview of all discovered prop drillings
and allows user navigation to anyone of them
through an interactive command line menu.
Once a prop drilling is selected,
the user has the option to proceed to refactoring
through any applicable strategy;
Component Composition, Context API or both.

\subsection{Evaluation Results}

\subsubsection{RQ1. How common are Prop Drilling occurrences in real-world \React{} applications?}

The purpose of RQ1 is to investigate
developer adoption of \PropDrilling{} in a realistic setting.
We answer RQ1 through analyzing all benchmark projects
for identification of \PropDrilling{} instances and
their inherent properties.
To analyze each codebase,
we have integrated into \ReactRefactor{}
advanced static analysis capabilities
for recovery of \React{} components and their properties from source code
and construction of the Component Hierarchy Graph.

Table~\ref{tbl:pd-occurrences} presents the results
on the prevalence of \PropDrilling{} instances across benchmark projects.
Column 2 provides the number of identified prop drillings (PDs),
while Column 3 presents the number of pure prop drillings (Pure PDs)
-- comprising exclusively pure prop forwardings -- and their ratio
over all prop drillings in parentheses.
In order to gain a more thorough insight on \PropDrilling{} adoption,
we have, also, estimated the number of \React{} components 
that serve as prop drilling participants,
as well as their ratio 
over the total number of identified \React{} components.
Column 4 presents the number of identified components for each codebase
and Column 5 provides the number of participants
and the respective ratio inside parentheses.

\begingroup
\setlength{\tabcolsep}{5pt} 
\renewcommand{\arraystretch}{1.3} 
\begin{table}[htb]
\centering
 \scriptsize
\begin{tabular}{l r r r r}
\multirow[b]{2}{0.6cm}{\theading{Project}} &
\multirow[b]{2}{0.6cm}{\theading{PDs}} &
\multirow[b]{2}{1.4cm}{\theading{Pure PDs}} &
\multicolumn{2}{c}{\theading{React Components}} \\
\cmidrule{4-5}
& & &
\multicolumn{1}{c}{\theading{All}} &
\multicolumn{1}{c}{\theading{\makecell{Prop Drilling \\ Participants}}} \\
\hline

grafana         & 1376 	& \valpct{ 684}{49.7} 	&       2728 	& \valpct{ 931}{34.1} \\
metabase        & 1917 	& \valpct{ 842}{43.9} 	&       2472 	& \valpct{1207}{48.8} \\
affine          &  841 	& \valpct{ 548}{65.2} 	&       1690 	& \valpct{ 669}{39.6} \\
superset        &  256 	& \valpct{ 146}{57.0} 	&        525 	& \valpct{ 136}{25.9} \\
novu            & 1334 	& \valpct{ 654}{49.0} 	&       1833 	& \valpct{ 742}{40.5} \\
lobe-chat       &  141 	& \valpct{ 134}{95.0} 	&       1185 	& \valpct{ 196}{16.5} \\
plane           &   14 	& \valpct{   6}{42.9} 	&       1358 	& \valpct{  23}{\ 1.7} \\
Rocket.Chat     &  430 	& \valpct{ 154}{35.8} 	&       1391 	& \valpct{ 425}{30.6} \\
cal.com         &   48 	& \valpct{  15}{31.3} 	&        559 	& \valpct{  50}{\ 8.9} \\
mastodon        &  219 	& \valpct{ 102}{46.6} 	&        393 	& \valpct{ 146}{37.2} \\
joplin          &   15 	& \valpct{   6}{40.0} 	&        122 	& \valpct{  19}{15.6} \\
redash          &   80 	& \valpct{  23}{28.7} 	&        252 	& \valpct{  87}{34.5} \\
mitmproxy       &   29 	& \valpct{   9}{31.0} 	&        116 	& \valpct{  23}{19.8} \\
prometheus      &   25 	& \valpct{  22}{88.0} 	&         59 	& \valpct{  25}{42.4} \\
carbon          &   69 	& \valpct{  58}{84.1} 	&        108 	& \valpct{  47}{43.5} \\
\cmidrule{1-5} 
\textbf{Total} 	& 6794 	& \valpct{3403}{50.1} 	&      14791 	& \valpct{4726}{32.0} \\
\cmidrule{1-5}

\end{tabular}

\captionsetup{justification=centering}
\caption{Prevalence of \PropDrilling{} usage in benchmark projects.}
\label{tbl:pd-occurrences}
\end{table}
\endgroup

A first remark on Table~\ref{tbl:pd-occurrences} results
is that several \PropDrilling{} instances
are identified in both small (<20KLOC) and large projects.
Specifically, 10 out of 15 benchmark projects
include at least 50 prop drillings,
while three of the largest projects 
(\proj{grafana}, \proj{metabase}, \proj{novu}) 
are characterized by more than 1000 occurrences.
Pure PDs account for 28--95\% of prop drillings per project
with a median value of 46.6\%,
indicating that data propagation is frequently 
the primary purpose of \PropDrilling{}.
Regarding \React{} components that are coupled
through at least one prop drilling,
they amount to at least 25\% of all identified components
in 10 out of 15 projects (median 34.1\%).
\proj{plane} represents an outlier in our dataset,
characterized by a low number of prop drillings 
and, consequently, a low participants' ratio (1.4\%) 
with respect to its codebase size.
The reason, revealed through code inspection,
is primarily the modular structure of \proj{plane},
comprising several libraries packaged as separate sub-projects
in the same Github repository.
Thus, component hierarchies span different sub-projects
and prop drillings cannot be effectively identified.
Overall, the results for RQ1 demonstrate 
the prevalence of \PropDrilling{} in \React{} codebases
and highlight the challenges faced by developers
in maintenance taks related to data propagation across component hierarchies,
reinforcing, thus, the applicability of the proposed refactoring. 

\begin{answer}
\PropDrilling{} is frequently encountered in real-world applications,
irrespective of their size.
The introduced coupling affects a considerable share of application components,
with prop drilling participants accounting for at least 25\% of all components
in several benchmark projects.
\end{answer}

\subsubsection{RQ2. What is the range of Prop Drilling complexity?}

In RQ2 we study the complexity of \PropDrilling{} instances
to understand the extent of coupling introduced among components
and its implications on maintainability.
To answer RQ2 we approximate the complexity of prop drillings
through the \emph{depth} and \emph{breadth} metrics.
Depth refers to the number of intermediate components
and captures the scope of the introduced coupling for data propagation.
Breadth represents the number of properties
simultaneously forwarded along the same path in a Compound Prop Drilling (CPD)
and indicates the coupling strength among participant components.

\begin{figure*}[!htb] 
    \centering
    \includegraphics[scale=0.43]{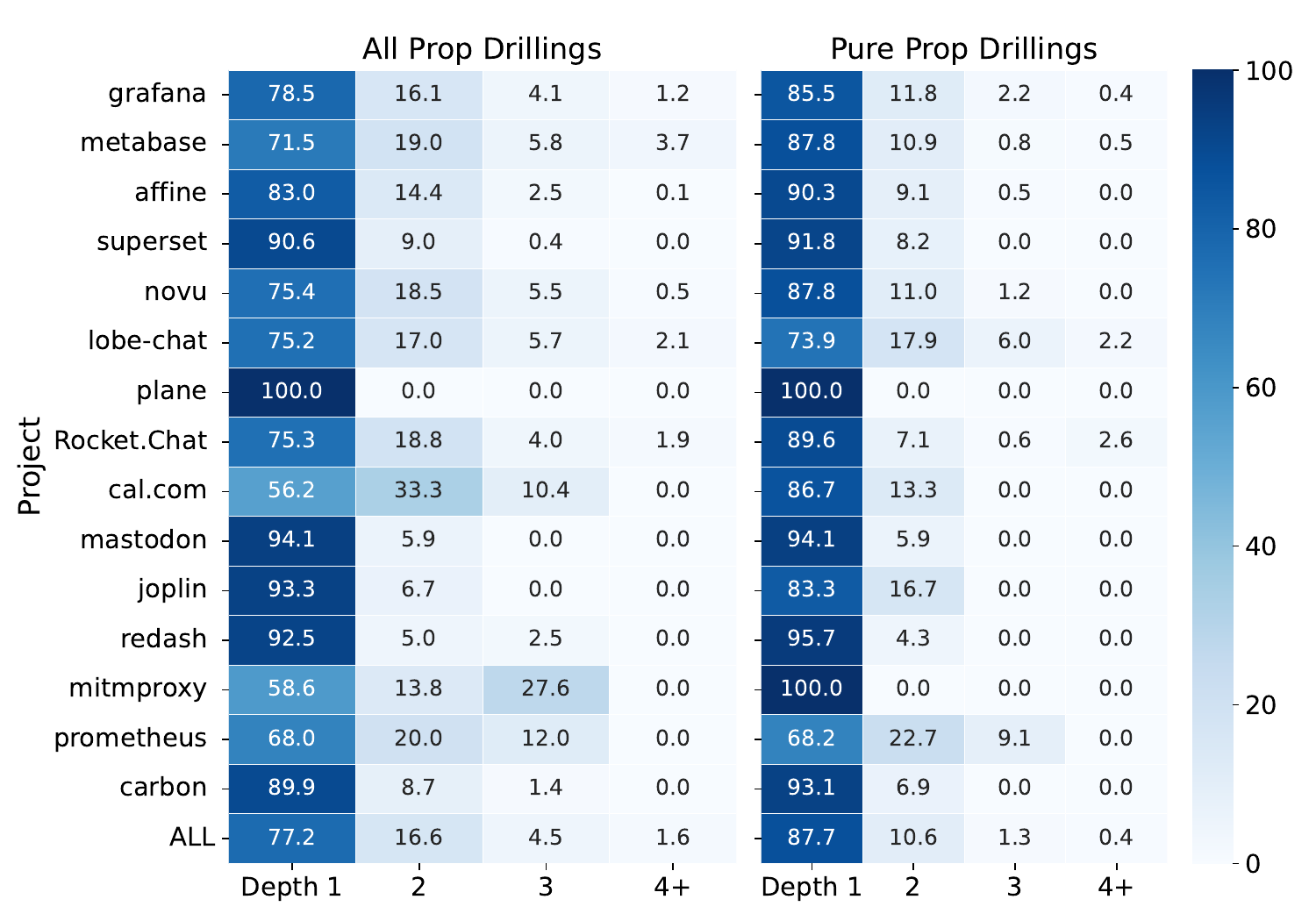}
    \captionsetup{justification=centering}
    \caption{Depth distribution for All Prop Drillings and Pure Prop Drillings.}
    \label{fig:depth-distr}
\end{figure*}

Figure~\ref{fig:depth-distr} presents results on \PropDrilling{} depth,
as extracted from the analysis of all benchmark projects.
The figure comprises two heatmaps,
each one illustrating the distribution of a set of prop drillings (PDs)
to depth levels 1--3 and four or more (4+).
The heatmap on the left presents the distribution of all identified PDs,
providing as values the percentage of PDs having a specific depth level.
The heatmap on the right focuses on pure prop drillings (PPDs)
and each value corresponds to a ratio over total PPDs for each project.
The last row ``All'' provides aggregate results across all projects.

Results on depth distribution show 
that a large share of PDs (77.2\%) and PPDs (87.7\%)
are characterized by the minimum depth (Depth 1).
Despite their relatively low complexity,
their frequent occurrence across projects
incurs a cumulative effect on component coupling,
due to several components acting as intermediates.
Furthermore, 
a considerable share of PDs (16.6\%) and PPDs (10.6\%) have a depth of 2.
Especially in large projects, 
these figures correspond in absolute terms to several PD instances,
e.g. more than 100 PDs and 50 PPDs 
in \proj{grafana}, \proj{metabase}, \proj{affine} and \proj{novu}.
Prop drillings with a depth of 3 
account for more than 5\% of all PDs in 6 projects,
whereas PPDs with the same depth exceed this threshold in only 2 projects.
Cases with a depth of 4 or greater are less frequent
and are concentrated primarily in \proj{grafana} and \proj{metabase}.

\begingroup
\setlength{\tabcolsep}{4pt} 
\renewcommand{\arraystretch}{1.3} 
\begin{table*}[htb]
\centering
 \scriptsize
\begin{tabular}{l r r r r r r}
\multirow[b]{2}{0.5cm}{\theading{Project}} &
\multirow[b]{2}{0.6cm}{\theading{CPDs}} &
\multicolumn{3}{c}{\theading{Breadth distribution}} &
\multicolumn{2}{c}{\theading{Member PDs}} \\

\cmidrule(lr){3-5} \cmidrule(l){6-7} 
& &
\multicolumn{1}{c}{\theading{2 (\%)}} &
\multicolumn{1}{c}{\theading{3} (\%)} &
\multicolumn{1}{c}{\theading{4+} (\%)} & 
\multicolumn{1}{c}{\theading{All} (\%)} &
\multicolumn{1}{c}{\theading{ \breadth{CPD} $\geq$3} (\%)} \\
\hline

grafana         &  254 & \valpct{ 158}{62.2}  & \valpct{  42}{16.5}  & \valpct{  54}{21.3}  & \valpct{ 753}{54.7}  & \valpct{ 437}{31.8}  \\
metabase        &  403 & \valpct{ 243}{60.3}  & \valpct{  96}{23.8}  & \valpct{  64}{15.9}  & \valpct{1112}{58.0}  & \valpct{ 626}{32.7}  \\
affine          &  166 & \valpct{ 123}{74.1}  & \valpct{  31}{18.7}  & \valpct{  12}{7.2}   & \valpct{ 397}{47.2}  & \valpct{ 151}{18.0}  \\
superset        &   57 & \valpct{  24}{42.1}  & \valpct{  23}{40.4}  & \valpct{  10}{17.5}  & \valpct{ 187}{73.0}  & \valpct{ 139}{54.3}  \\
novu            &  266 & \valpct{ 184}{69.2}  & \valpct{  46}{17.3}  & \valpct{  36}{13.5}  & \valpct{ 724}{54.3}  & \valpct{ 356}{26.7}  \\
lobe-chat       &   21 & \valpct{  17}{81.0}  & \valpct{   2}{\ 9.5} & \valpct{2}{\ 9.5}    & \valpct{  48}{34.0}  & \valpct{  14}{\ 9.9}   \\
plane           &    3 & \valpct{   2}{66.7}  & \valpct{   1}{33.3}  & \emptyVal{}          & \valpct{   7}{50.0}  & \valpct{   3}{21.4}  \\
Rocket.Chat     &   91 & \valpct{  65}{71.4}  & \valpct{  20}{22.0}  & \valpct{   6}{\ 6.6} & \valpct{ 219}{50.9}  & \valpct{  89}{20.7}  \\
cal.com         &   10 & \valpct{   7}{70.0}  & \valpct{   2}{20.0}  & \valpct{   1}{10.0}  & \valpct{  24}{50.0}  & \valpct{  10}{20.8}  \\
mastodon        &   58 & \valpct{  49}{84.5}  & \valpct{   4}{\ 6.9} & \valpct{   5}{\ 8.6} & \valpct{ 132}{60.3}  & \valpct{  34}{15.5}  \\
joplin          &    1 & \emptyVal{}          & \emptyVal{}          & \valpct{   1}{\,100} & \valpct{   5}{33.3}  & \valpct{   5}{33.3}  \\
redash          &   17 & \valpct{  13}{76.5}  & \emptyVal{}          & \valpct{   4}{23.5}  & \valpct{  42}{52.5}  & \valpct{  16}{20.0}  \\
mitmproxy       &   11 & \valpct{  11}{\,100} & \emptyVal{}          & \emptyVal{}          & \valpct{  22}{75.9}  & \emptyVal{}          \\
prometheus      &    4 & \emptyVal{}          & \valpct{   3}{75.0}  & \valpct{   1}{25.0}  & \valpct{  15}{60.0}  & \valpct{  15}{60.0}  \\
carbon          &   16 & \valpct{   8}{50.0}  & \valpct{   6}{37.5}  & \valpct{   2}{12.5}  & \valpct{  44}{63.8}  & \valpct{  28}{40.6}  \\
\cmidrule{1-7} 
\textbf{Total} 	& 1378 & \valpct{ 904}{65.6}  & \valpct{ 276}{20.0}  & \valpct{ 198}{14.4}  & \valpct{3731}{54.9}  & \valpct{1923}{28.3}  \\
\cmidrule{1-7}
\end{tabular}

\captionsetup{justification=centering}
\caption{Compound Prop Drillings and their breadth distribution.}
\label{tbl:pd-compound-distr}
\end{table*}
\endgroup

Table~\ref{tbl:pd-compound-distr} presents results 
on the frequency of Compound Prop Drillings (CPDs) across benchmark projects
and their distribution to various \emph{breadth} levels.
Column 2 provides the number of identified CPDs per project,
while Columns 3--5 present the distribution of CPDs 
to breadth levels 2, 3 and $\geq$4, respectively.
In Columns 4 and 5 we focus on \emph{Member PDs},
i.e., PDs that are constituent parts of CPDs.
Column 4 presents the number of PDs 
that serve as members of at least one CPD
and provides in parentheses 
their ratio over the total PDs of each project.
Finally, Column 5 focuses on PDs 
that are members of CPDs with breadth
greater or equal to 3 (\breadth{CPD}$\geq$3).

Based on the results in Table~\ref{tbl:pd-compound-distr}
we observe that CPDs are prevalent across benchmark projects.
Their member PDs (Column 6) account for more than half (54\%) of all identified PDs,
with the proportion ranging from 33\% to 75\% across projects.
Thus, more than half of the forwarded properties 
are propagated alongside at least one other property in the same path,
increasing the coupling among PD participants.
More specifically, property pairs are propagated in 65.6\% of all CPDs,
since they are characterized by the minimum breadth (Column 3),
while another 20\% of all CPDs forward 3 properties (Column 4).
CPDs of breadth 4 or greater are, also, common among the largest projects
with \proj{grafana}, \proj{metabase} and \proj{novu} 
including 54, 64 and 36 instances, respectively (Column 5).
Overall, CPDs exceeding the minimum breadth account for 34.4\% of all CPDs
and their member PDs represent a notable share (28.3\%) of all PDs (Column 7).
These findings indicate the potential negative impact of CPDs on code complexity
and motivate the development of tools for their automated detection
and elimination, particularly in large projects.

\begin{answer}
Prop drillings (PDs) generally exhibit low depth complexity.
Higher-depth instances ($\geq 2$) account for about 20\% of all PDs,
a non-negligible share that corresponds to several occurrences
particularly in large projects.
In contrast, breadth complexity is more prevalent,
as 54.9\% of all PDs participate in Compound Prop Drillings (CPDs).
Moreover, 28.3\% of all PDs are members of CPDs 
with increased breadth complexity ($\geq 3$).
\end{answer}

\subsubsection{RQ3. What ratio of Prop Drillings can be eliminated 
through fully automated refactoring?}

In RQ3 we investigate the feasibility of automated refactoring for identified PDs
and the applicability of the \emph{Context API} 
and \emph{Component Composition} refactoring strategies.
To answer RQ3, 
we evaluate each PD instance 
against two sets of refactoring preconditions, one for each strategy.
Depending on the preconditions satisfied,
a PD instance is classified as candidate for refactoring
through the Context API, Component Composition
or both strategies.
The evaluation and classification process is
automated by the \ReactRefactor{} tool.

\begin{figure*}[!htb] 
    \centering
    \includegraphics[scale=0.44]{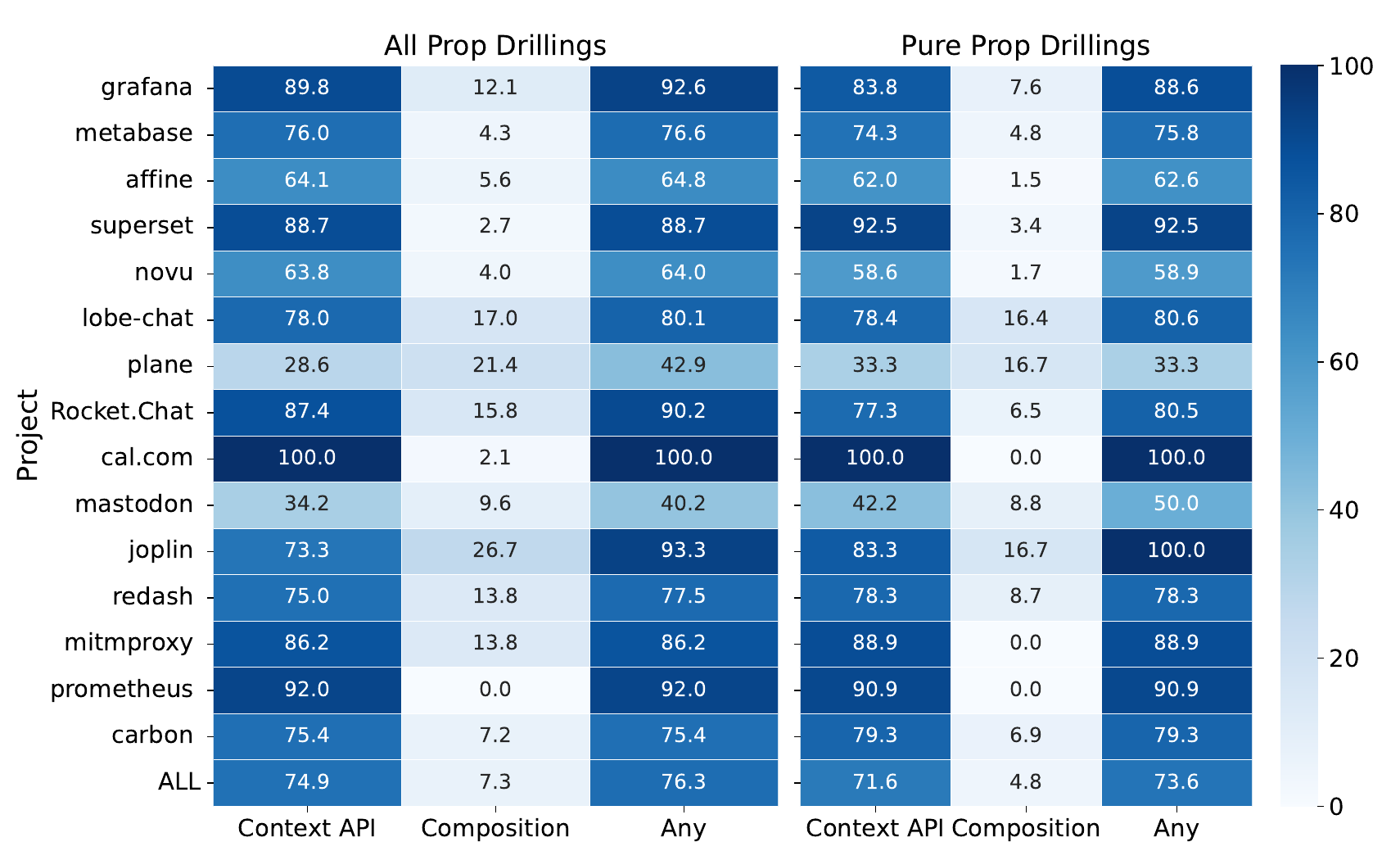}
    \captionsetup{justification=centering}
    \caption{Applicability of refactoring strategies to  All Prop Drillings and Pure Prop Drillings.} 
    \label{fig:ref-distr}
\end{figure*}

Figure~\ref{fig:ref-distr} presents results
on the identified refactoring candidates among PDs and PPDs,
as well as on the applicability of each refactoring strategy.
The figure includes two heatmaps focusing on PDs and PPDs, respectively.
The heatmap on the left presents for each benchmark project
the share of PDs that can be refactored through
the Context API, Component Composition or any strategy.
The latter, practically, represents 
the percentage of refactoring candidates among PDs.
The heatmap on the right provides the respective ratios
with respect to the total PPDs of each project.
Finally, the last row ``All'' provides aggregate results across all projects.

The results in Figure~\ref{fig:ref-distr} highlight the broad applicability 
of the Context API refactoring strategy.
Specifically, the strategy enables the automated refactoring 
of 74.9\% of all identified PDs and 71.6\% of PPDs.
At the project level, it is applicable to
at least 28.6\% of PDs (median value 76\%) and
33.3\% of PPDs (median value 78.4\%). 
On the other hand, refactoring candidates with Component Composition
account for 7.3\% of all PD instances and 4.8\% of PPDs.
As expected, Component Composition has more limited applicability
due to its stricter preconditions compared to the Context API strategy.
Overall, 76.3\% of all PDs and 73.6\% of PPDs 
can be refactored using at least one of the considered strategies.
The latter account for 36.8\% of all PDs 
and represent higher-impact refactoring opportunities,
as they target cases dedicated solely to data propagation
and completely eliminate forwarded properties from intermediate components.

\begin{answer}
Prop drillings (PDs) that can be eliminated through automated refactoring
account for 76.3\% of all PDs identified in the benchmark projects.
Refactoring through the Context API applies to 74.9\% of PDs
and Component Composition to 7.3\% of all PDs,
with some PDs being candidates for both strategies.
Refactorable pure prop drillings account for 36.8\% of all PDs
and represent a substantial set of of cases
with greater potential for code simplification.
\end{answer}

\section{Threats to validity}
\label{sect:ttv}

In this section we discuss the factors threatening
the validity of our empirical study.
We categorize and prioritize potential threats
according to the affected validity aspects,
based on the classification scheme suggested by~\citet{Wohlin2000}:
internal, external, construct and conclusion validity.
For each threat, we discuss the selected
action or strategy for its mitigation.

Threats to \emph{internal validity} refer to factors,
neglected or not controlled,
that influence the observed causality 
between treatment and outcome of an experiment.
Since our study is exploratory 
and does not seek to confirm a causality relationship,
the most relevant threat to internal validity,
among those documented by ~\citet{Wohlin2000},
is how unbiased is the selection of the \React{} applications
comprising our benchmark.
We mitigate this threat by constructing our benchmark dataset
from all projects used in relevant empirical studies
published in top-tier journals.

\emph{External validity} threats concern 
the generalization of empirical evaluation results.
A potential threat relates to the extent 
to which the benchmark dataset
is representative of the broader population of \React{} applications.
This threat is partly mitigated by the diversity of the benchmark projects
in terms of size, application domain and implementation language.
However, the dataset does not include closed-source projects,
potentially affecting the generalization of results to enterprise settings
characterized by different development practices and conventions.
Considering this threat, 
we decided to trade off this aspect of external validity 
in favor of the reproducibility and transparency of the empirical evaluation.

Threats to \emph{construct validity} concern the correspondence 
between theoretical constructs 
and the observations of the empirical evaluation.
A threat to construct validity is related to potential bugs
in the implementation of our research prototype,
that may lead to missed PDs or false positives.
We partly mitigate this threat 
through manual inspection of several identified PDs
and implementation of a test suite for \ReactRefactor{}
that verifies its basic capabilities.

\emph{Conclusion validity} concerns the statistical significance
of the results on causality relationships studied in an experiment. 
Since we conduct an exploratory study, 
where results and observations shape the findings
rather than confirm the validity of certain hypotheses, 
we evaluate \emph{reliability} as a counterpart of conclusion validity~\cite{Wohlin2000}.
Reliability concerns how dependent are the data and their analysis
to the specific researchers that conducted the study.
A potential threat to \emph{reliability} concerns the reproducibility of this study.
We alleviate this threat 
through releasing \ReactRefactor{} as open-source software
and publishing the required artifacts for study replication
as a Zenodo archive~\cite{reactRefactor26}.

\section{Conclusions -- Future Work}
\label{sect:conclusions}

We have proposed a method
for identifying \PropDrilling{} instances in a \React{} codebase
and eliminating them using two automated refactoring strategies
based on the \emph{Context API} and \emph{Component Composition},
native \React{} capabilities 
that facilitate the implementation of adaptable and reusable components.
Building on a formal specification of the \PropDrilling{} concept,
our method employs static analysis 
to recover a graph representation of the \React{} codebase --
the Component Hierarchy Graph (CHG) -- 
which captures parent-child dependencies among components
and the flow of data due to prop passings and prop forwardings.
Furthermore, the proposed method introduces 
a CHG traversal algorithm for identifying \PropDrilling{} instances
and a set of refactoring preconditions 
that determine whether a \PropDrilling{} instance is refactorable 
and which refactoring strategies are applicable.
Finally, we specify source code transformation procedures
for eliminating a \PropDrilling{} instance through either refactoring strategy.
The proposed method has been implemented
as a Node.js command line tool, \ReactRefactor{}, 
and empirically evaluated on a benchmark dataset of 15 open-source \React{} projects.
The results of the empirical study revealed
the frequent occurrence of prop drillings in real-world applications,
irrespective of their size.
Regarding \PropDrilling{} complexity,
the \emph{depth} of identified instances is generally low;
nevertheless, higher-depth instances ($\geq 2$) 
account for about 20\% of all identified prop drillings
and occur particularly in large projects.
\emph{Breadth} complexity is more prevalent,
as 54.9\% of all prop drillings participate in Compound Prop Drillings.
A notable finding of our study is 
that a substantial share of all prop drillings (76.3\%) are refactorable,
with the Context API strategy applying to 74.9\% of all instances
and Component Composition to 7.3\%.
Overall, the results demonstrate the potential of the proposed method
to identify data propagation complexity
and mitigate it through native \React{} capabilities.

Future extensions of this work 
could investigate additional refactoring strategies
to increase the range of prop drillings eligible to automated refactoring.
Moreover, in-depth study of other forms of tangled prop drillings 
-- beyond Compound Prop Drillings -- 
could lead to refactoring techniques 
that alleviate broader data propagation issues.
Finally, an interesting direction for future work 
is to investigate the generalization of the proposed method 
to other component-based UI frameworks,
such as Vue.js or Android Jetpack Compose,
which exhibit analogous data propagation patterns.
This could provide further insights 
into the complexity of data propagation 
across component-based UI technologies.




\bibliographystyle{elsarticle-num-names}
\bibliography{bibliography}

\end{document}